# Low Temperature Plasma for Biology, Hygiene, and Medicine: Perspective and Roadmap

Mounir Laroussi[1], *Fellow, IEEE,* Sander Bekeschus[2], Michael Keidar[3], Annemie Bogaerts[4], Alexander Fridman[5], XinPei Lu[6], Kostya (Ken) Ostrikov[7], Masaru Hori[8], Katharina Stapelmann[9], Vandana Miller[10], Stephan Reuter[11], Christophe Laux[12], Ali Mesbah[13], James Walsh[14], Chunqi Jiang[15], Selma Mededovic Thagard[16], Hiromasa Tanaka[8], DaWei Liu[6], Dayun Yan[3], Maksudbek Yusupov[4]

[1]Electrical & Computer Engineering Department, Old Dominion University, Norfolk, VA, 23529 USA. Email: mlarouss@odu.edu
[2]ZIK Plasmatis, Leibnitz Institute for Plasma Science and Technology (INP), 17489 Greifswald, Germany
[3] Mechanical and Aerospace Engineering Department, The George Washington University, Washington, DC, 20052 USA
[4]Research Group PLASMANT, Department of Chemistry, University of Antwerp, Antwerp, Belgium
[5]Nyheim Plasma Institute, Drexel University, Philadelphia, PA, USA
[6]School of Electrical and Electronics Engineering, Huazhong University of Science and Technology, Wuhan, Hubei, P.R. China
[7]School of Chemistry and Physics, Queensland University of Technology, Brisbane, QLD 4000, Australia
[8]Center for Low-temperature Plasma Sciences, Nagoya University, ES 425, Furo-Cho, Chikusa-Ku, Nagoya 464-8693, Japan
[9]Department of Nuclear Engineering, North Carolina State University, Raleigh, NC, USA
[10]Department of Microbiology and Immunology, Institute of Molecular Medicine and Infectious Disease, Drexel University College of Medicine, Philadelphia, PA, USA
[11]Department of Engineering Physics, Polytechnique Montreal, Montreal, QC, Canada
[12]Laboratoire EM2C, CNRS, CentraleSupelec, Universite Paris Saclay, Paris, France
[13]Department of Chemical and Biomolecular Engineering, University of California, Berkeley, CA 94720, USA
[14]Department of Electrical Engineering and Electronics, University of Liverpool, Brownlow Hill, Liverpool, UK L69 3GJ
[15]Frank Reidy Research Center for Bioelectrics, Old Dominion University, Norfolk, VA, USA
[16]Department of Chemical and Biomolecular Engineering, Clarkson University, Potsdam, NY 13676 USA

## Abstract

Plasma, the fourth and most pervasive state of matter in the visible universe, is a fascinating medium that is connected to the beginning of our universe itself. Man-made plasmas are at the core of many technological advances that include the fabrication of semiconductor devices, which enabled the modern computer and communication revolutions. The introduction of low temperature, atmospheric pressure plasmas to the biomedical field has ushered a new revolution in the healthcare arena that promises to introduce plasma-based therapies to combat some thorny and long-standing medical challenges. This paper presents an overview of where research is at today and discusses innovative concepts and approaches to overcome present challenges and take the field to the next level. It is written by a team of experts who took an in-depth look at the various biomedical applications, made critical analysis, and proposed ideas and concepts that should help the research community focus their efforts on clear and practical steps necessary to keep the field advancing for decades to come.

## CONTENTS

## I. INTRODUCTION

M. Laroussi, A. Bogaerts, and XP Lu

The multidisciplinary field of the biomedical applications of atmospheric pressure low temperature plasma (AP-LTP) has been 25 years in the making [1 - 12]. Since its introduction in the mid-1990s the field has expanded tremendously and has reached several critical and very important milestones, which include clinical trials on wound healing and cancer treatment. These advancements and successes have been accomplished thanks to the outstanding contributions of many investigators and research groups from around the world. A quarter century may seem like a long time for an individual investigator but for a typical field of science it is a relatively short period of time. Yet, the advances and accumulation of new knowledge about the interaction of plasma with living soft matter has been nothing short of impressive. Today, our scientific knowledge concerning the physical and biochemical pathways involved in the interaction of plasma with biological cells and tissues has reached a good level of maturity. However, more remains to be done both on the fundamental and application levels. So, 25 years seems like a good time marker for the research community to take a good look and evaluate the current state of the field, identify the challenges and obstacles, and propose ideas on how these could be overcome. These are the main objectives of this roadmap paper. A team of experts in the various sub-topics of the field has been tasked to each summarize the present status of their sub-topic, inform the research community about the remaining challenges, and propose potential approaches and methods that can be employed to overcome these challenges.

LTP produces a rich "cocktail" of reactive species including reactive oxygen species, ROS, and reactive nitrogen species, RNS, which have been shown to play pivotal roles in the observed biomedical outcomes [13]. LTP sources also exhibit high electric fields and produce photons at various wavelengths/energies, which can also play synergistic biological roles. The state of knowledge regarding the effects of these plasma-produced agents is quite advanced as of the present, but many of the effects on the subcellular and molecular levels are either not fully understood or a matter of debate. To help elucidate these effects, advanced modeling and simulation approaches have been brought to bear on many of these issues [14] – [17]. In addition, in the aim of making LTP biological effects reproducible under similar operating conditions, it is important to develop control methods that allow for the minimization of the effects of environmental conditions, such as humidity, temperature, air pressure (e.g., altitude), power fluctuations, etc. while at the same time optimizing the treatment as a function of the properties of the target (shape, surface morphology, thickness/depth, conductivity, etc.). These are some of the outstanding issues that are discussed in this paper. Ideas and new concepts to tackle these challenges are proposed, providing an outlook on what still needs to be done and where the field is headed in the near future.

The paper is organized in seven sections. The first section discusses low temperature plasma sources. The application of plasma in biology and medicine would not have been possible without the development of atmospheric pressure low temperature plasma sources. However, there are still outstanding issues, such as control of these sources to achieve optimal performance for any application they are used for. The second section discusses the application of LTP in biological decontamination. This was the first bio-application of LTP and the initial seed that was at the genesis of the entire field. This application has proved to be of the utmost importance, because of the advent of antibiotic resistance of bacteria and pathogenic proteins (such as prions), and because of viral-driven pandemics, such as the present COVID-19 pandemic. The third section discusses the use of LTP to treat water and air. Plasma treatment of water is used for nitrogen fixation, reduction of pollutants, and inactivation of pathogens in the food cycle, food safety, and for water purification. For air treatment, traditionally plasma has been used for the inactivation of airborne pathogens. For example, HVAC systems can be retrofitted with a plasma treatment stage to purify indoor air from biological contaminants. However, with the advent of viral-driven pandemics (such as COVID-19), plasma treatment of air has taken a renewed importance. The fourth section discusses LTP-assisted wound healing. This was the first patient-centered medical application of LTP and maybe one of the most developed and has undergone extensive animal studies followed by clinical trials at various hospitals, especially in Germany. One of the therapeutic challenges is the healing of chronic wounds. Chronic wounds, such as diabetic ulcers, do not heal easily or at all. Thousands of amputations occur every year in the US alone, because of the inability of conventional methods to satisfactorily deal with this problem. LTP has been shown to safely coagulate blood and enhance the proliferation of cells such as fibroblasts, which made gas plasma an attractive technology for wound healing. The fifth section deals with the application of LTP in the field of dentistry. Here LTP has been used in three areas of dentistry: periodontal, endodontic and prosthodontic treatments. LTP has also been used "indirectly" for surface modifications of implants, oral surgery to improve interfacial bonding strength of dental objects used in periodontics or prosthodontics, and sterilization of dental instruments. The sixth section discusses cancer treatment by LTP, what is referred to as "plasma oncology". This application has seen great progress in the last decade that recently culminated in preliminary clinical trials in a few hospitals in Germany and the USA. *In vitro* Investigations have shown that LTP can kill cancer cells and *In vivo* experiments have shown that LTP treatment can shrink tumors size and/or delay their growth. However, there is still much work to be done to understand the effects of LTP on cancer cells on the sub-cellular and molecular levels and the effects on the tissues surrounding tumors. Finally, the seventh section of this paper discusses modeling and simulation works that have been conducted so far to further elucidate the plasma-cell and plasma-tissue interactions. The interaction of LTP with entire cells or tissue can up to now only be simulated on a macroscopic scale, obviously providing less detailed information than modeling the interactions with individual cell components, such as the phospholipid bilayer, DNA or proteins, which is possible on the molecular level. Much progress has been made in recent years, but more work has to be done, and major challenges are the compromise between calculation time and level of detail of the simulations, the need for accurate input data and experimental validation, and the translation of the obtained insights towards improving the biomedical applications.

To conclude, research on the biomedical applications of low temperature plasma has, first of all, produced new, previously missing, fundamental scientific knowledge regarding the interaction of the fourth state of matter with soft biological matter. Second, LTP is proving to be a technology that can help meet many of today's healthcare challenges, including the inactivation of antibiotic-resistant bacteria, the healing of chronic wounds, treatment of various dermatological diseases, and even some types of cancers. More recently, agricultural applications have also emerged, which include treatment/decontamination of fruits and legumes for longer shelf life, the modification of the wettability of the surfaces of seeds, and the enhancement of the germination speed and yield of plant seeds. Finally, LTP may also be a viable technology for space medicine. Here plasma can replace perishable medication and can be used to decontaminate tools and gear from possible extraterrestrial microorganisms and to disinfect wounds and/or skin infections that astronauts may experience during long-duration space missions. To ensure that LTP technology can meet all the requirements of effectiveness and safety for the above listed applications, the research community needs to identify all the remaining scientific and technical challenges and seek creative solutions that can take the field to the next level. We hope that this paper plays a role in this endeavor and provides some guidance for future investigations.

## II.  Low Temperature Plasma Sources

S. Reuter, A. Mesbah, D. Liu, X. Lu, M. Laroussi, and K. Ostrikov

### II.1 Current state

#### Plasma Sources:

Two main types of low temperature plasma (LTP) sources have been used in plasma medicine research, dielectric barrier discharges (DBDs) and non-equilibrium atmospheric pressure plasma jets (N-APPJs) [1], of which many varieties exist. N-APPJs quickly became widely applied LTP sources because they can generate a stable and controllable plasma volume outside the confinement area of electrodes within the surrounding environment [2]. This is crucial for a direct plasma treatment of diseased tissue on a patient [3]. Biologic effects of LTP sources in medicine include sterilization/decontamination of biotic and abiotic surfaces, the promotion of growth factors, control of cell migration, induction of apoptosis, angiogenesis, microcirculation and modulation of the immune response [4] – [6]. These effects have been attributed to the reactive oxygen and nitrogen species (RONS), electric field, charged particles and ultraviolet radiation (UVR) generated by LTP [7 - 9]. Because LTP can help preventing the aerosol transmission of respiratory infectious diseases such as tuberculosis and the Coronavirus disease 2019 (COVID-19) [10, 11-13], recent research focus was set on plasma sources for reduction of pathogenic biological aerosols (PBAs).

A second classification of biomedical LTPs is application mode oriented – two key types are distinguished: direct and indirect plasma sources [10]. For the first type, the plasma has direct contact to the treated surface, the second type does not. In the case of direct LTP sources, the treatment target (skin, wound, tissue, etc.) acts as the counter electrode, therefore the discharge current (displacement or conduction current) flows through the body. One example of direct LTP sources is shown in Figure II.1(a) [4,14]. A typical plasma jet is shown in Figure II.1 (b). A pin electrode covered by a dielectric tube with one end closed is the power electrode [15] while the ring electrode outside the tube is the ground electrode [16]. A commercial version of this type of N-APPJ is the kINPen [17]. More configurations of plasma jets can be found e.g., in [2] and [15].

For indirect LTP sources, the plasma is mainly produced between the electrodes and some reactive species of the plasma can reach the surface of the target via the electric field propagation (ionization wave), convection and diffusion mechanisms [15]. An air DBD-based indirect LTP source is shown in Figure II.1(c) [18]. A commercial version of this type of LTP is the PlasmaDerm device. Convection through the air flow delivers RONS generated by the plasma inside the tube to the treatment target. Another indirect LTP, the surface dielectric barrier discharge (SDBD) shown in Figure II.1 (d) A thin plasma layer forms between the grounded electrodes' structures. The reactive species and UV light can interact with the object surface due to the short distance between the grounded electrode and the target's surface [5].

Recently, air plasma sources which can generate stable air discharges under conditions of high air flow rate have been used to sterilize PBAs. Figure II.1 (e) shows the structure of bioaerosol disinfector based on an atmospheric-pressure positive corona discharge. Another example of DBD plasmas in air that can inactivate aerosolized *Bacillus subtilis* cells and *Pseudomonas fluorescens* indoor and outdoor is shown in Figure II.1 (f). [19, 20]. It is important to note that the large plasma volume covers almost the entire cross section of the quartz tube, largely increasing the interaction intensity between the plasma and PBAs.

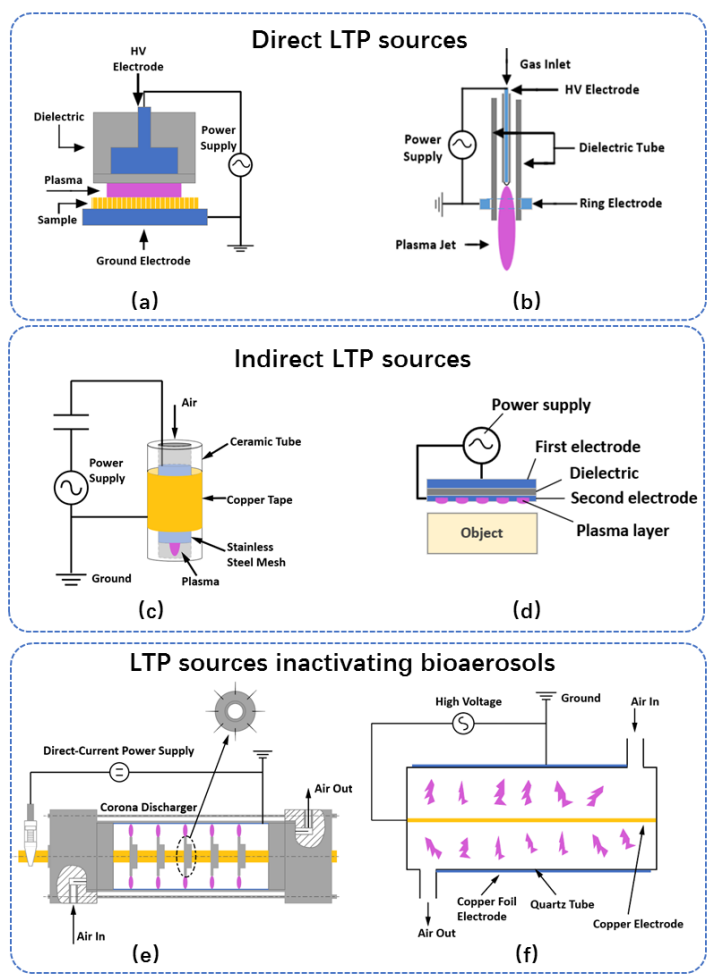

Fig. II.1 Schematics of low temperature plasma (LTP) sources. Direct LTP sources: (a) The volume-air dielectric barrier discharges (DBDs) with a gas gap in the mm range. (b) Typical DBD plasma jets. Indirect LTP sources: (c) Air DBD was generated inside the ceramic tube. (d) Surface air DBDs. LTP sources inactivating bioaerosols: (e) The needle corona discharge array to inactivate Penicillium aerosols. (f) The air DBD system (quartz tube) to kill airborne bacteria or virus.

*Plasma Processes:*

In the past decade, enormous progress has been made in the characterization of atmospheric pressure LTP sources. This can be attributed to the technology development of diagnostics, as well as to a deeper knowledge of plasma chemistry and processes through the combination of modelling and diagnostics. Through reference plasma sources [21] and a continuing development of high-precision diagnostic methods for absorption spectroscopy [22], emission and laser spectroscopy [23], and non-optical methods such as mass spectrometry [24], a thorough understanding of the plasmas has been achieved [9, 23, 25, 26].

Most recent developments to understand plasma processes focus on electric fields in plasmas, electron properties and energy dissipation pathways [27]. The emergence of ultrafast diagnostic techniques has enabled researchers to observe the generation of reactive species and space resolved electric field dynamics in plasma breakdown relevant time scales [28]. Finally, especially for the medical field, the interaction and transport processes of the plasma generated reactive species into liquid environments are fundamental [29]. Substantial progress has been made in the ability to control plasma activated solutions and using mixed-phase systems such as plasma-bubble systems to enhance the reactivity and other relevant physical and chemical progress. These advances have been made through the close interaction of physicists, engineers and chemists.

*Plasma Control:*

Control over plasma composition is mandatory for a targeted medical application. Tremendous progress has been made in controlling electron dynamics e.g., through a multi-frequency approach [30]. Ensuring a reproducible composition by using a tailored gas surrounding the active effluent has opened the door to several promising fundamental biomedical studies. Traditionally, control of LTPs has primarily relied on statistical process control strategies, as widely used in semiconductor manufacturing. However, such control strategies are inherently open-loop, in which the lack of on-line sensing and feedback corrective action can compromise the reliability and repeatability of LTPs due to the intrinsic plasma variabilities and external disturbances. The key challenges in feedback control of LTPs stem from 1) the nonlinear interactions between multiple input variables (i.e., discharge parameters) and multiple output variables of a plasma discharge (i.e., plasma effects), 2) the need to constrain the plasma properties to ensure safe plasma treatment and circumvent adverse plasma-induced surface effects, and 3) the need to realize the proper synergy between the chemical, physical, and electrical effects of LTPs when interacting with complex biological surfaces [31]. Recently predictive control and learning-based control approaches have shown exceptional promise for safe, effective and reproducible operation of N-APPJs [32,33,34]. Reduced-order, physics-based models and data-driven models have been used to develop model predictive control (MPC) strategies for constrained, multivariable control of the nonlinear effects of plasma on surfaces [32,35]. Experimental studies using cell and chemical assays have correlated operational parameters of CAPs such that real-time monitoring of the plasma's chemical effects can be effectively utilized in an adaptive manner for feedback control [36]. These studies have been extended to develop MPC-based strategies for targeted cancer cell treatment using real-time electro-chemical impedance spectroscopy measurements [37]. Furthermore, deep neural network-based controllers have been demonstrated to be particularly useful for control of N-APPJs using resource-limited embedded systems, which are especially suitable for portable and point-of-use biomedical applications [38].

*II.2 Challenges and proposed solutions*

Challenges

Despite the fact that many LTP sources suitable for biomedical applications have been developed and successfully applied *in vitro* and *in vivo*, the number of clinical trials remains relatively small [39]. One of the crucial challenges relates to the controlled delivery of plasma agents to diseased targets. The challenges are similar to those of plasma processes in the semiconductor industry in the 1990s that arose due to the significant lack of knowledge about the fundamental mechanisms involved in the plasma-materials interactions. A concerted R&D effort in academia and industry allowed the identification of the plasma parameters and substrate properties and geometries that led to the desired manufacturing process. Dry etching by plasma thus became commonplace technology in semiconductor production. The plasma medicine community is currently at a similar development stage with lessons to be learned from the semiconductor processing experience which necessitated substantial numerical modelling assisted process analysis [40 - 42].

A stark difference between the technological and the medical application of LTPs is the certification procedures of medical plasma sources. While more and more plasmas are used in medical research and some countries have certified plasma sources for medical use, a clear path to bringing a new type of plasma source to clinical application remains to be found.

Although more progress in the field of new LTP sources and innovative biological applications are desirable, there are fundamentals issues that still need to be addressed. These can be subdivided into the three areas *Plasma physics and chemistry*, *biomedicine*, and *LTP technology development*. The following questions require further investigations to be able to reach a knowledge-based plasma medicine approach:

*Plasma physics and chemistry*:

How do electric field dynamics in N-APPJs lead to the generation of energetic electrons, ions and reactive oxygen and nitrogen species? What is the role of rotational-vibrational-translational energy transfer? How can the concentrations and fluxes of RONS be accurately controlled? How can plasma induced reactivity be delivered through different media without undue influence of environmental factors?

*Biomedicine*:

What are the effects of treatment target properties and composition (skin, the wound covered by effusion, tumors, aerosols, etc.) on LTP characteristics? How do we standardize the flux and energy of LTP transmitted to cells and tissues to appropriately define "plasma dose"?

*LTP technology development*:

How can we use artificial intelligence (AI), new materials, and new power electronics technology to optimize the design of the LTP sources and control the plasma? How do we ensure that LTP biomedical applications are patient compatible? How can LTP technology be designed sustainably to enable end-of-life re-cycling and re-use of the electrodes, electronic components and other parts of the plasma devices? How can plasma sources be adapted to the respective treatment requirements and be safely applicable *in-vivo* at the tumor site also in non-surgical procedures?

<u>Proposed Solutions</u>

   LTP sources development is an interdisciplinary research area, therefore, collaborations between plasma physics and chemistry, power electronics, fluid dynamics, microbiology and medical science are required (Figure 2). In order to set up the roadmap for applied physics here, we concentrate on the challenges in the three areas mentioned above.

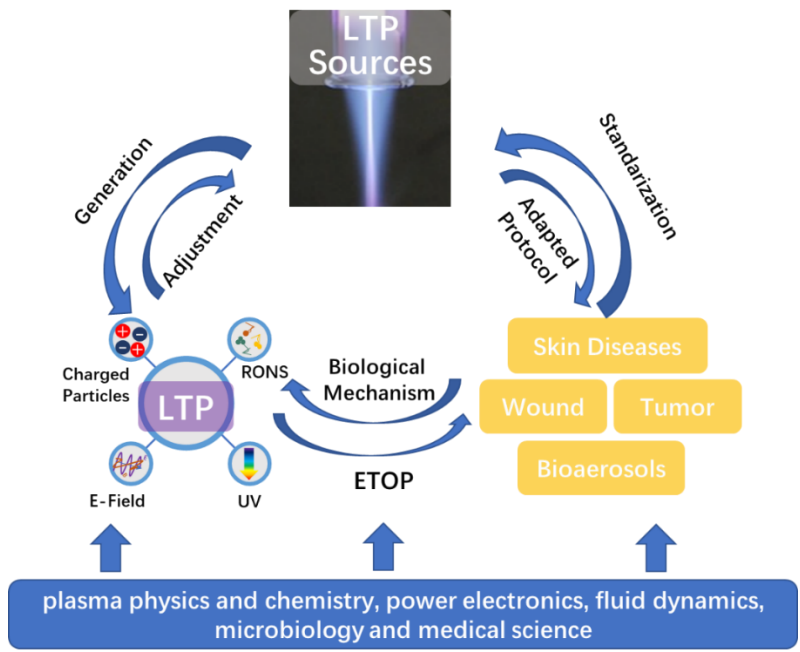

Fig. II.2 Three key areas affecting the development of LTP sources – 1) plasma physics and chemistry 2) Biomedicine and 3) LTP technology development. The collaboration of multidisciplinary scientific community will enhance the understanding of fundamental mechanisms of operation of LTP sources. Feedback loops between the key areas include reactive species generation and delivery, protocols and adapted protocols as well as the equivalent total oxidation potential (ETOP).

   It is necessary to carry out experimental and simulation work to thoroughly elucidate and determine all the physical and chemical processes in a cohesive way. For example, in the case of N-APPJs, using fast imagery with high spatial resolution (μm range) and high time resolution (sub-ns range) along with 2D or 3D simulation models, the effect of photo-ionization, seed electrons and electric field can be obtained.

   To understand where and how the active constituents generated by the plasma, reactive oxygen and nitrogen species (RONS), UV, charged particles, and electric fields are generated, their interactions, and interdependence is important. Measurement techniques such as infrared absorption spectroscopy for radical and molecule detection, optical emission spectroscopy for discharge dynamics, Stark spectroscopy or electric field induced second harmonic generation for electric field, Thomson, Raman and Rayleigh scattering for electron parameters, laser induced fluorescence (LIF) for radical and molecules detection, and mass spectrometry can be employed. 2D or 3D simulations with hundreds of plasma reactions will facilitate the identification of the main production mechanisms of the plasma reactive species (see chapter on modelling).

   When LTP sources are used to treat fluid-covered wounds or bioaerosols, plasma comes in contact with liquids. Therefore, knowledge of the liquid-based chemistry will improve the understanding of the discharge processes happening at the gas–liquid interface. The evaporation of liquid is an important process, because it not only affects the plasma chemistry, but also decreases the lifetime of bioaerosols. On the other hand, when LTP is used to treat skin diseases such as psoriasis, the scurf or accretive tissue may decrease the delivery of plasma, making it necessary to adjust the plasma treatment dose.

Since the reactive components of the plasma consist of a large variety of species it is difficult to standardize plasma treatments. A proposed metric is the equivalent total oxidation potential (ETOP) (See Figure 2) which includes all RONS and UV/VUV generated by the plasma, could be used to evaluate the plasma dose [43]. If the ETOP of a plasma source is known at different values of power, temperature and humidity, this could facilitate the expansion and application of LTP technology in medicine.

Plasma source development ranges from small to large scale devices and the design is driven by the target application. In any case, moving towards mobile and personalized devices will greatly enhance application fields to include ambulant, point of care or even in a home care environment. Battery powered devices or low power devices are promoting this development (e.g. [44]).

The rapid development of artificial intelligence (AI) can be used to transform the mathematical modeling, real-time diagnostics, and optimal operation of LTP sources for applications in plasma medicine [45]. In addition, new oxidation-resistant and high-strength electrode materials as well as new power semiconductor devices can enhance the efficiency of LTP sources and reduce the cost of their development and manufacturing. Environmental considerations should be taken into consideration to reduce emissions of harmful gases and meet personnel minimum exposure requirements of gases such as ozone.

Finally, the certification issues need to be addressed. The market is steadily growing for plasma medicine, but a standardized certification procedure is still missing, while first attempts have been made for a minimal characterization requirement [46]. Lacking certification procedures could ultimately pose a great risk to the field, due to insufficiently safe plasma sources entering the market.

## III. LTP for Biological decontamination

J. Walsh, M. Laroussi, M. Keidar, and C. Laux

### III.1 Current State

The use of low-temperature plasma for the decontamination of biological agents has been under intensive investigation for several decades. Commercial LTP decontamination systems, employing low-pressure plasma, are nowadays widely used in healthcare settings across the world for the sterilization of temperature sensitive medical devices. While low pressure LTP systems are undoubtedly efficient for biological decontamination, the added cost and complexity of a vacuum based process limits their broader appeal. At the end of the twentieth century, interest in the use of plasma for biological decontamination was reignited with a demonstration that LTP generated under ambient pressure conditions could be harnessed for the non-thermal inactivation of microorganisms [1]. Much of this early work focused on the use of noble gas plasma jets due to their ease of use and convenience [2], with more efficient and inevitably more complex LTP-based decontamination devices being developed over the subsequent years to meet a growing number of application needs.

As the area of research continues to expand, it is becoming more diverse and varied. While early works focused primarily on the use of LTP to eliminate planktonic bacteria, recent efforts have focused on more representative contamination such as antibiotic resistant bacteria [3], mixed species biofilms [4], and endospores [5]; conditions that present a formidable challenge to any non-thermal decontamination technology. Beyond bacterial contamination, the global SARS-CoV-2 pandemic has placed a spotlight firmly on the need to develop fast, efficient and low-cost viral decontamination techniques that can be easily deployed beyond a healthcare setting. Given the minimal equipment costs and the possibility of consumable-free operation of LTP, many members of the plasma science community have turned their attention to viral decontamination. These efforts have resulted in a number of published studies within the past 12 months that comprehensively demonstrate the efficiency of LTP against viral agents [6], [7], a clear indication that the plasma science community is both flexible to adapt and willing to tackle some of the greatest challenges facing humanity.

*III.2 Challenges and Proposed Solutions*

<u>Current and Future Challenges</u>

Underpinning the use of LTP for biological decontamination is the multiphase interaction between a non-equilibrium gas plasma and a target organism; given the shear complexity of this interaction a plethora of scientific challenges remain unresolved. These challenges essentially fall into distinct areas, such as the need to understand and control fundamental LTP processes, and the need to understand the biological effects of LTP exposure at the target, Figure 1. Adding to the complexity is the two-way interaction between the LTP and its target, often giving rise to emergent and synergistic behaviors that are difficult to predict, characterize and control.

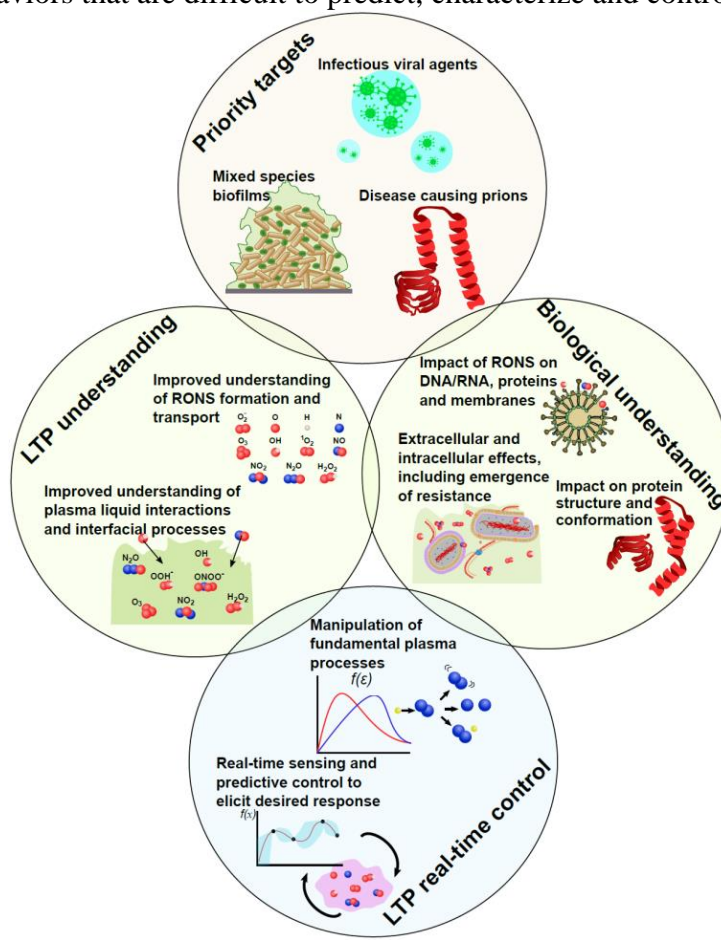

Fig.III.1 Diagram showing the priority decontamination targets for LTP and challenges relating to the understanding and control of LTP for effective, efficient, and repeatable biological decontamination.

Perhaps one of the greatest contemporary challenges associated with LTP mediated decontamination relates to unravelling the underpinning mode of action that determines the efficacy of the process; a challenge that is further exasperated by the large number of different plasma systems currently under investigation, with each system presenting different operating characteristics and parameters, combined with the vast number of biological agents being explored [5]. Great progress has been made in this area over the past decade, with increasingly sophisticated techniques being adopted to gain new levels of insight. Despite this progress, more needs to be done, especially at the biomolecular level. Critically, understanding the mode of action is only the first step in developing an effective LTP decontamination system. Being able to manipulate and control the fundamental LTP processes to elicit a desired biological response at the target remains the 'holy grail' in the area, yet such control extends beyond our current capabilities. A key advantage of LTP technology is its ability to be applied directly at the point of need, in many situations this is likely to be beyond the laboratory in an environment that is in a constant state of flux. Given that much of the reactivity of an LTP system can be derived from its interactions with the surrounding environment, be it from air entrainment within a plasma jet plume [8], or through the direct generation of LTP in ambient air [9], the question of how to generate a consistent discharge composition in a constantly changing environment remains unresolved.

While understanding and controlling LTP processes is the key to ensure efficiency and repeatability of the decontamination processes, many challenges relating to the underpinning biomolecular processes must also be addressed. Typically, these challenges are highly specific to the biological target:

*Bacteria and biofilm formation*:

In the natural environment, microorganisms cluster together and form a protective layer as a means of self-preservation, this community is known as a biofilm. Typically, microorganisms encased within a biofilm are 1000 times more resistant to antimicrobial agents than their planktonic counterparts. When LTP interacts with a biofilm the antimicrobial agents created must first penetrate this protective layer, known as the Extracellular Polymeric Substances (EPS). The complex composition of the EPS acts to inhibits the transport of reactive agents to the encased microorganisms, significantly diminishing decontamination efficacy. Biofilms also present ideal conditions for the formation of antimicrobial resistance, one of the greatest challenges facing humanity in the 21$^{st}$ century. Indeed, the key questions of if, how and how quickly microorganisms form resistance to LTP remain unanswered.

*Viruses*:

The use of LTP to decontaminate viruses has been under active investigation for several years, yet it was not until the emergence of SARS-CoV-2 and the subsequent pandemic that the approach gained mainstream attention. Recent approaches have demonstrated effective LTP mediated surface and airborne SARS-CoV-2 decontamination [7]. More now needs to be done to understand the underpinning mode of antiviral action and demonstrate efficacy and repeatability. Compounding this challenge is our ability to test LTP systems under real-world conditions; SARS-CoV-2, like many other viruses of concern, are highly infectious and any scientific study must be conducted in a laboratory with an appropriate containment level. Many researchers do not have access to such facilities and have therefore adopted surrogate models, such as the use of bacteriophages and pseudoviruses [10], [11]. Despite extremely promising results, the applicability of these models remains unclear, and more work is needed to understand how the model results translate to real-world efficacy. Another challenge in thus area relates to how best to apply LTP, with it potentially being used to remove airborne viruses that are in an aerosolized form or applied directly to decontaminate surfaces.

*Infectious proteins*:

Prions are misfolded proteins with the ability to transmit their misfolded shape onto normal variants of the same protein. Prion diseases are a family of rare progressive neurodegenerative disorders that affect both humans and animals. Typically, prion contaminated materials are extremely difficult to decontaminate, with reports indicating prions remain infective following incineration at 600ºC [12]. Consequently, prions present a formidable challenge for LTP decontamination yet numerous studies have reported promising results [13]. Several challenges are yet to be resolved in this promising area, perhaps the most important of which relates to the underpinning mode of action and impact of LTP exposure on the protein structure and conformation. Another key issue that is yet to be fully addressed relates to the fate of the exposed prions, are they completely eradicated or are they simply dislodged from the surface by LTP and transported elsewhere?

Proposed Solutions

   The LTP community has an ever-growing arsenal of sophisticated diagnostics techniques at their disposal; methods such as Laser Thomson Scattering (LTS) and femtosecond Two-photon Laser Induced Fluorescence (fs-TALIF) are becoming more widespread and can provide an unprecedented level of insight into fundamental plasma processes [14], [15]. Despite this progress, the shear level of complexity and multiscale nature of plasma-bio interactions dictates that experimental techniques alone will not be sufficient to uncover a full understanding; computational modelling therefore has a major role to play. While computational plasma models are becoming increasingly advanced and provide a deeper level of insight than ever before [16], the introduction of a realistic biological entity into such models extends beyond our current ability. To overcome this formidable challenge, a fusion of different modelling techniques will be essential, with approaches such as those used in molecular dynamics being integrated within conventional LTP modelling codes to provide a new insight at the biomolecular level.

   In addition to our increasing knowledge of LTP processes, more insight into the biomolecular processes induced by LTP exposure is essential to guide process optimisation and instil confidence in potential end-users. In terms of virus inactivation, understanding the impact of LTP generated components on viral DNA/RNA alongside proteins and membranes is essential. Techniques such as RNA/DNA sequencing should be adopted to characterise the viral genome, monitor viral gene expression and genome replication in response to LTP exposure. While the efficiency of LTP mediated virus inactivation should be assessed on appropriate cell lines using different standardised methods, such as the plaque assay and the immunohistochemical focus forming assay [17]. To further enhance our knowledge of LTP mediated biofilm denomination genomic, proteomic and metabolomic studies are vital to obtain a quantitative understanding of the spatial and temporal variation in the microenvironment of cells embedded within the biofilm EPS and how LTP-introduced agents affect cell-cell communication. Antimicrobial resistance remains a key challenge for any novel decontamination technology and the impact of LTP exposure on the emergence of resistance and adaptation of microorganisms remains unclear; addressing this will require dedicated genotypic approaches such as PCR and DNA hybridization methods. Finally, LTP interaction with infectious proteins requires a detailed investigation into possible changes to the protein structure and conformation, using both classical techniques such as X-ray crystallography and Nuclear Magnetic Resonance (NMR), coupled with emerging techniques such as cryo-electron microscopy (cryo-EM) [18].

   Armed with a clear understanding of the fundamental processes at play the question of LTP manipulation and control must be addressed. Given that it is desirable to operate LTP under ambient pressure conditions, the resulting plasma is highly collisional hence many of the techniques successfully developed to control low-pressure plasma are ineffective. Advanced power sources employing dual frequency excitation or short voltage pulses (nanosecond and sub-nanosecond duration) may offer the exciting ability to manipulate the electron energy distribution function high-pressure LTP and therefore provide a means to manipulate RONS generation. While such techniques are promising, knowing how to control LTP processes is only one part of the puzzle, knowing how to control them to achieve the desired biological response in an ever-changing environment is the ultimate key to successes. Addressing this challenge will undoubtedly require a closed-loop control methodology, informed by real-time sensing and diagnostics. Very few LTP systems currently under investigation employ feedback and are thus highly susceptible to fluctuations in the external environment, ultimately impacting their efficiency and repeatability [19], [20]. To support the development of real-time and closed loop control systems many lessons can be drawn from the field of data science, machine learning and artificial intelligence. Exciting preliminary results have demonstrated that predictive control methodologies can be applied to harness LTP [21], with such strategies facilitating the control of a complex black-box function, such as an LTP interacting with a biological target, with little prior knowledge of the systems characteristics.

To conclude, as the development of LTP technology for biological decontamination spans the field of plasma science, engineering, and the life sciences; consequently, a multidisciplinary approach is therefore essential to make progress. While the LTP community has been extremely successful in demonstrating the potential of the technology for biological decontamination in a plethora of exciting application areas, further engagement with scientists from beyond the LTP domain is essential to overcome the formidable research challenges that remain. As with other LTP applications in the biomedical field, a key challenge is understanding the complex underpinning mechanisms of plasma action, an endeavour that can only succeed through a cohesive and multidisciplinary approach.

## IV.  PLASMA TREATED AIR AND PLASMA TREATED WATER

K. Stapelmann, S. Mededovic Thagard, and A. Fridman

### IV.1 Current state

The growing use of atmospheric pressure plasma opened the door for their applications in the treatment of water and air. During the last decade, plasma treatment of water has been investigated for nitrogen fixation, reduction of pollutants, and inactivation of pathogens in the food cycle, food safety, and water purification arenas. The treatment of air, which traditionally has been focused on the inactivation of airborne pathogens in air flows and on their chemical cleaning, has gained significant renewed interest due to the COVID-19 pandemic

*Plasma treated water for the food cycle and food safety:*

Low-temperature plasma (LTP) science and technology has found its place in agriculture for applications ranging from sustainable fertilizer production to food safety and reduction of pollution and pathogens. Sustainable farming is an international priority due to world population growth, global warming and decreases in arable land. Sustainable farming requires efficient management of water and nitrogen-based fertilizers to conserve scarce water resources, minimize environmental impact and reduce fossil-fuels now used for fertilizer production, mainly with the Haber-Bosch process. To supplement the global nitrogen demand and offer environmental protection a more flexible process can be developed with plasma-treated water (PTW), see Figure IV.1 for an example PTW reactor. PTW has been used for the irrigation of seeds and seedlings, showing a growth-promoting effect.  This opens up the possibility of precisely manipulating plant productivity while enhancing plant resilience to biotic and abiotic stresses [1]. PTW for fertigation (fertigation is injection of fertilizers and chemicals into an irrigation system) of plants can be used to supplement nitrogen in soil and soilless systems, as well as to control pH and bacterial load in recirculating nutrient solutions in soilless systems. In case of seed irrigation with PTW, the seeds are exposed to hydrogen peroxide ($H_2O_2$) and nitric oxides ($NO_x$), which may trigger changes in plant hormones, promoting germination. LTP offers the unique opportunity to tailor and control reactive species produced.

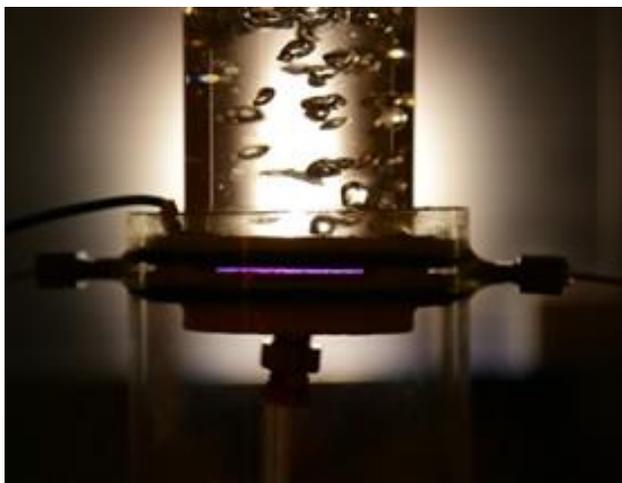

Fig. IV. 1 A dielectric barrier discharge for nitrogen fixation in water. The plasma is in direct contact with the water surface. To increase the residence time and the surface area to volume ratio, air is flown through the plasma and bubbled into a 500 ml water column. (courtesy of Plasma for Life Sciences Laboratory, North Carolina State University).

Addressing another grand challenge, the reduction of post-harvest loss, PTW has been applied to fresh produce to reduce microorganisms and potential chemical contaminants on food [2]. It is well known that plasma and PTW have antimicrobial effects [3–5] and expanding the knowledge from plasma sterilization and plasma medicine into the plasma agriculture sector is a logical step.

PTW can be generated in bulk liquids and applied as a sanitizer [6,7]. It has been shown that PTW retains or even promotes food quality while effectively inactivating microorganisms with minimal impact on chemical, nutritional, texture, and sensory attributes of food produce [2,8]. Significant water conservation in fresh produce washing and protective treatment has been achieved by replacing bulk wash water with plasma-activated micron- and submicron- water mist. In addition to significant decrease of wash water consumption, the plasma-mist processing of fresh produce demonstrated high efficacy of bacterial inactivation [9].

The efficient generation and transport of reactive species into the liquid is the key to advancing the widespread use of plasma devices for microbial inactivation, nitrogen fixation and for treating different source water, as discussed next. Regardless of the application, the process owes its strong chemical conversion/oxidation and disinfection capabilities to the generation of reactive oxygen and nitrogen species (RONS) in the gas phase and their subsequent solvation into the liquid phase and also has the advantage of synergistic effects of high electric fields, and UV/VUV light emissions. The best performing plasma reactors feature a high surface area to volume ratio which ensures increased solvation rates [10]. Plasma in contact with a liquid on large surfaces, plasma in bubbles in liquids, and water droplets in the plasma have all been utilized to increase this ratio [11].

*Purification of water supplies:*

The first report on utilizing electrical discharge plasmas to directly degrade organic contaminants dates back to the 1980s when Clements *et al.* reported decolorizing a dye by using an electrical discharge within a liquid [15]. Since then, over a hundred different plasma reactors have been developed to treat a range of contaminants of environmental importance including pharmaceuticals, herbicides, pesticides, warfare agents, bacteria, yeasts and viruses using direct-in-liquid discharges with and without bubbles, and discharges in a gas over and contacting the surface of a liquid giving rise to electrohydraulic reactors, bubble reactors, spray reactors, hybrid gas-liquid reactors, falling water film reactors, etc. Different excitation sources including AC, nanosecond pulsed and DC voltages have been utilized to produce pulsed corona, corona-like, spark, arc, and glow discharges, among other discharge types. Plasma-based water treatment (PWT) generates a range of RONS and other species, as explained in the previous section, requires no chemicals, and can be optimized for batch and continuous processing. The best performing plasma reactors effectively utilize plasma-generated species by concentrating the contaminants at the plasma-liquid interface and, as noted earlier, feature high surface area to volume ratios. However, despite the obvious benefits and advantages of PWT, the technology is only now reaching a level of development where it can be commercially used to treat emerging and legacy contaminants, as demonstrated recently on pharmaceutical waste [16] and groundwater contaminated with poly- and perfluoroalkyl substances (PFAS), Figure IV.2 [17].

PWT is generally considered a highly energy intensive process and a significantly more complex technology compared to the existing advanced oxidation processes. However, the scientific evidence to support these claims is scarce and often misleading as the plasma-based treatment of PFAS has been shown to outperform all existing destructive technologies in terms of energy efficiency and treatment time on bench and pilot scales by an order of magnitude [18]. To overcome the widespread skepticism towards plasma as an advanced oxidation technology and position the process at the forefront of water purification, the technology cannot be simply as good as the other oxidation processes but significantly better and able to treat a wide range of common water constituents.

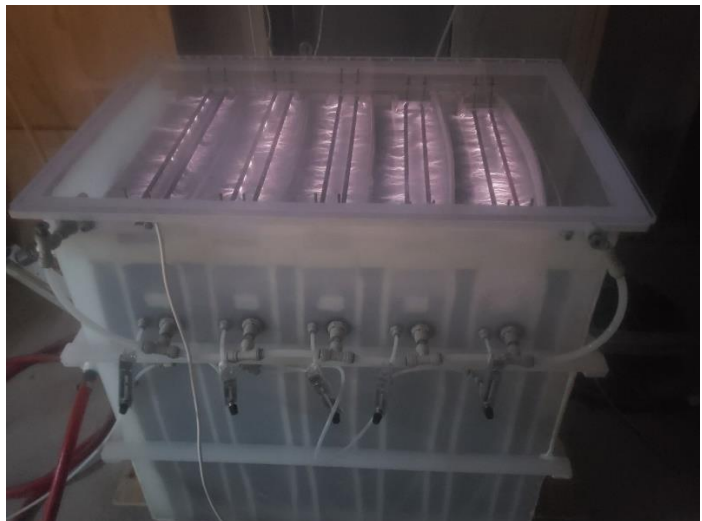

Fig IV.2. A commercial 10 gpm water treatment system based on electrical discharge plasma (courtesy of Plasma Research Laboratory, Clarkson University).

*Decontamination of air supply:*

Another use of plasma attracting significant interest is the treatment of air focused both on inactivation of airborne pathogens in air flows, as well as on their chemical cleaning. The recent coronavirus related problems, environmental protection challenges, as well as the growing threat of bioterrorism have brought into focus the need for effective methods to decontaminate and sterilize not only different surfaces, but also gases and especially air. In the early 2000s, most of the plasma-based air sterilizers have been successful only when coupling the plasma technology with high efficiency particulate air (HEPA) filters to both trap and kill microorganisms [20]. Usually, such coupling of filtration and plasma systems means trapping the micro-organisms with the filter followed by sterilization of the filter using non-thermal plasma discharges.

A few studies, some inspired by the COVID-19 pandemic and recently reviewed by Mohamed *et al.* [21], have shown the efficacy of plasmas in inactivating airborne pathogens without the use of a filter specially to prevent virus transmission. Most commonly, dielectric barrier discharges (DBD) are used in different configurations, see Figure IV.3 for an example. DBDs offer the advantages of operating in ambient air and at room temperature. As mode of action, the short-lived reactive species OH and singlet oxygen have been suggested. Hence, the most promising results reported in literature were obtained when aerosols were passed through active plasma volume (rather than effluent).  The challenge of air treatment in ventilation systems is the short residence time on the order of subseconds of the air inside the active plasma region.

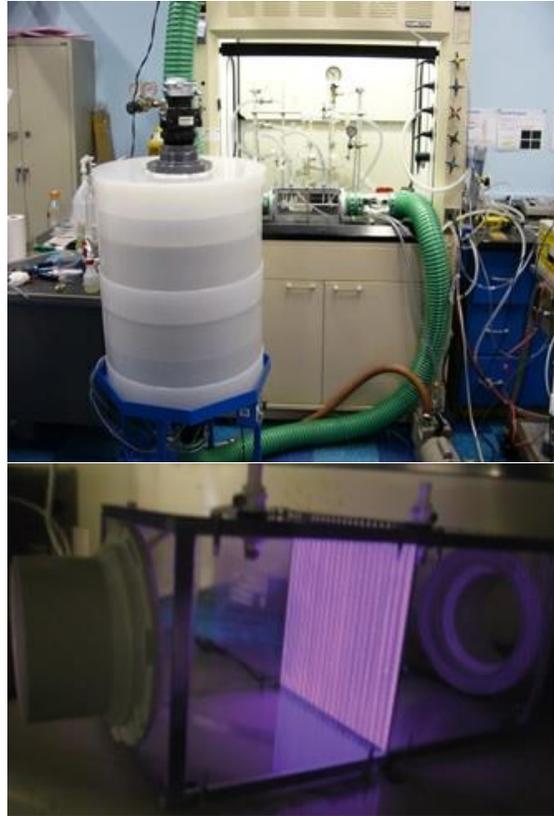

Fig IV.3 Pathogen Detection and Remediation Facility (PDRF): above - general view, below - Dielectric Barrier Grating Discharge (DBGD) with air sterilization chamber. [23]

Plasma air treatment, similar to plasma water treatment, is generally considered a highly energy intensive process and a significantly more complex technology compared to the existing air cleaning technologies. However, the current gold standard for the reduction of airborne pathogens is filtering them out with HEPA filters. These filters effectively retain particles, such as bacteria, viruses, and pollen without inactivating them. The filters cause significant pressure losses in heating, ventilation, and air conditioning (HVAC) systems giving rise to higher energy and maintenance costs. The HVAC system itself has to be designed for the pressure losses which does not allow for a later upgrading of the HVAC system with such filters without substantial costs. Especially during the COVID-19 pandemic, air cleaning systems that can be installed in existing HVAC systems would have been of great advantage. As the *Decadal Assessment of Plasma Science* conducted by the National Academies of Science, Engineering, and Medicine stated in May 2020: "…we may have missed an opportunity to have another tool at our disposal to aid in the current health crisis" [22].

*IV.2 Challenges and Proposed Solutions*

<u>Present and Future Challenges</u>

*Plasma treated water for the food cycle and food safety:*

To date, most of the literature available using PTW for fertigation focuses on plant growth in early growth stages, typically in soilless systems, as recently reviewed by Ranieri *et al.* [1]. The effect of PTW on plants needs to be studied beyond the seedling stage for the process to become a viable alternative for agriculture. Further, knowledge about the effects of PTW on plant growth in actual soil systems and the impact of PTW on soil health and the microbiome structure is limited. The effect of other RONS generated in PTW beyond $NO_3^-$ is also not well understood. Further research is needed on $H_2O_2$ and NO which may act as signaling molecules and were proposed as strategies to enhance crop stress adaptation and to prolong shelf life [12]. To understand the impact of PTW on the whole plant and its ecosystem, multi-disciplinary research approaches including plant biologists, soil scientists, microbiologists, and plasma scientists and engineers are necessary.

For the translation of laboratory results to field conditions, the impact of organic matter in water on the plasma chemistry also needs to be considered and studied. Tap water or well water is typically used for irrigation in greenhouses and fields. It is well known that the buffering capacity of tap water influences the liquid chemistry [11,13]. While the use of deionized or ultrapure water in the lab is justified for the reproducibility of results and to gain fundamental understanding in simpler systems, the presence of organic matter that may become part of chemical reaction pathways [14] needs to be studied in more detail to understand, optimize, and eventually commercialize PTW for nitrogen fixation.

The use of PTW or plasma-activated mist in food safety applications is a promising field. However, limited information is available about the impact on produce on a molecular scale beyond pH, color and texture. Some studies report discoloration or decrease in firmness after exposure to PTW, some reported no reduction in terms of vitamins while others reported increased vitamin D content, for example. The scientific community is using a variety of different plasma sources, with different excitation mechanisms, gas compositions, different liquid volumes and inconsistent times between PTW preparation and use. Due to the transient nature of many reactive species believed to be crucial for inactivation of bacteria, the time between PTW preparation and application is of significant importance, yet not always reported in the literature. Given the varieties of produce and plasma systems used, it is difficult to understand the fundamental science of PTW interacting with fresh produce.

*Purification of water supplies:*

Despite the convoluted effects of a multitude of parameters affecting the performance of plasma reactors, there is increasing evidence that the current plasma reactors are superior for the treatment of surfactant compounds (e.g., PFAS, foams and dyes) but are unable to competitively treat non-surfactants that in reality constitute the majority of the compounds of environmental importance. Therefore, increasing the conversion efficiency for the treatment of non-surfactants to the levels of surfactant-like compounds, and increasing the energy efficiency and throughput for the treatment of all compounds represent and remain the key challenges for PWT.

The slow progress in the development and scaleup of plasma reactors in general is rooted in three fundamental challenges. The first one relates to the lack of understanding of the basic processes in the plasma, in the bulk liquid and at the plasma-liquid interface, including the nature of their interaction. The second challenge surrounds the plasma reactor design and the lack of understanding of the extent to which classical chemical reactor design theory is applicable to the development of (heterogeneous) plasma reactor systems. There also exists a large gap between the fundamental discoveries and translation of those discoveries into engineering solutions. The plasma reactor scaleup and the applicable scaling laws belong to the third category of challenges.

The plasma reactor development for water purification purposes has been and still is largely trial and error, with research predominantly conducted in simple systems containing a single compound and using a range of different plasma excitation sources. Plasma diagnostics to quantify the plasma physics and chemistry remains a challenge. The bulk liquid composition and transport have been recognized as significant factors affecting the reactor performance and more research is needed using techniques such as laser-induced fluorescence (LIF) and particle image velocimetry (PIV) coupled with kinetic experiments to determine the relative importance of diffusive and convective mass transport and elucidate their individual roles in compound degradation [24,25]. For a range of organic compounds, we are still unable to differentiate among different plasma agents responsible for their degradation, which hinders the plasma reactor design and optimization. The roles of individual ROS, RNS, V(UV), ions, metastables and solvated electrons including that of the electric field are largely unknown due to the numerous issues associated with the existing techniques, protocols, assays, and commercial kits that have been developed. It is largely due to the absence of the adequate probe(s) and/or techniques to measure the concentrations of dissolved short-lived species that the community is still unable to learn how to control the plasma and bulk liquid chemistry with the input operational parameters and ultimately maximize the production of oxidative species or reductive species (plasma is a both an oxidative and reductive process) for a particular water treatment application.

To understand the fundamental science behind interfacial dynamics and develop accurate theories that describe process performance, more data is usually better. However, to upscale a plasma reactor by any degree requires information on only the key processes and design parameters that could be empirical. To date, ensuring a high plasma area to liquid volume ratio has been identified as the critical requirement but also the key technological challenge in the plasma reactor scaleup [19].

*Decontamination of air supply:*

Three major challenges are related to successful commercial and industrial application of plasma sterilization and cleaning of air flows: energy cost, productivity, and selectivity.

1. *Energy cost of plasma air treatment.* Plasma air decontamination systems consume electric energy, which determines most of the operational cost of the relevant technologies. The electric energy cost is a crucial challenge of plasma air cleaning since the majority of conventional and novel competing air cleaning technologies are less electric energy intensive (or do not consume electric energy at all). Cost is especially important for decontamination of large air volumes. Analysis of cost effectiveness shows that to be economically viable the electric energy consumption must be below $1\text{-}3 \times 10^{-3}$ kWh/m$^3$ of air for decontamination of large air volumes.

2. *Thruput (capacity) of the plasma system.* To be effectively commercialized, the air flow capacity of plasma systems should be scaled up to typical industrial/commercial levels of thousands of m$^3$ per hour. Even at the expected energy effective regimes (electric energy cost below $1\text{-}3 \times 10^{-3}$ kWh/m$^3$), the airflows on the level of thousands of m$^3$ per hour require at least 10 kW of cold plasma power. This is a significant engineering challenge that necessitates the development of powerful cold plasma discharges and building relevant power supplies with high efficiency matching characteristics.

3. *Selectivity of air cleaning processes.* Plasma systems may generate *non-acceptable treatment byproducts* like ozone, nitrogen oxides, etc. These byproducts restrict types and operational regimes of applied non-thermal plasma discharges. Detailed understanding of relevant plasmochemical process mechanisms and analysis of plasma system combination with other air treatment approaches are still needed.

<u>Proposed Solutions</u>

*Plasma treated water for the food cycle and food safety:*

As a common theme for plasma agriculture applications, upscaling is one of the biggest challenges. In order to optimize the process and to design highly efficient plasma-water or plasma-mist reactors, more fundamental understanding is required of a) the kinetics and collisional processes in the highly non-equilibrium plasma regime, b) the (efficient) transport of species into the liquid, and c) the plasma-generated species required for the optimum outcome. For a more fundamental understanding, advances in diagnostic capabilities in the liquid phase and at the plasma-liquid interface are required, as well as advances in modeling, capturing the properties of non-equilibrium plasma kinetics and the transport through the plasma-liquid interface into the liquid. The effect of charged species and photon-induced effects, as well as reactions involving short-lived species like atomic oxygen have yet to be addressed as many reaction rates, cross sections, and rate constants are unknown. Calculations and measurements of these unknown quantities are critical for advancing the understanding of the plasma-liquid interfacial processes.

Regarding the applications of plasma-treated water in the food cycle and food safety arenas, the plasma parameters affecting the enhanced plant growth and resilience, healthy soil and nutrient management must be better understood to optimize the process. While the effect of plasma-generated nitrate or ammonium as fertilizer is well known, other RONS and their potential benefits for plant growth and the plants' ecosystem have not been studied in detail yet. For food safety applications, the impact of RONS in the liquid on food quality and nutrients is also not well understood. The use of appropriate controls, e.g. nitrate, nitrite, and hydrogen peroxide controls with a concentration matching the concentrations in PTW, would help to identify mechanisms and to pave the road for upscaling and commercialization efforts.

*Purification of water supplies:*

One of the biggest challenges in the design, optimization and scaleup of highly efficient plasma reactors is finding the way for the community to work together towards this common goal and build on each other's findings. A large volume of literature currently exists on the effectiveness of over a hundred different plasma reactors for treating dozens of different compounds and biological species. And while each of these studies is unarguably useful and informative, systematic comparison of the compound(s)/microorganisms studied is needed, including their degradation/inactivation rates and starting concentrations, reactor types, discharge energies, excitation sources, and treatment volumes to be able to deconvolute and extract any information regarding the key design and operational principles that control the system performance. While such studies are rare due to their laborious nature, they are extremely informative and may set empirical guidelines for the plasma reactor design [19].

To remedy this issue and essentially calibrate multiple concurrent reactor design and scaleup research efforts, the community should agree on using and characterizing a reference plasma treatment system similar to the COST Reference Microplasma Jet that is currently being used in the biomedical community to identify the parameters and processes that guide and control the plasma reactor design and the process performance. An example of such a system would be a point gas discharge in a noble gas contacting the surface of the treated liquid with a grounded plate immersed in the liquid. Common electrode material, electrode dimensions, reactor volume, and the compound/microorganism treated would be chosen. Beyond these set parameters, immediate new research would welcome investigations on the excitation source types, process gas types, mixing, etc. and include advanced spectroscopic/laser techniques to quantify the plasma physics and chemistry. In the liquid, complementary analytical measurements would be focused on the quantification and/or identification of the extent of the compound degradation, byproducts and their toxicity, reaction mechanisms and the species responsible for the chemical transformations. Concurrently, techniques such as LIF and PIV would be applied to determine the location of degradation/inactivation, visualize the flow and measure the velocity fields in the liquid. The outcomes of these fundamental investigations would assist in determining the limitations and the advantages of the technology, its level of performance with respect to the existing technologies and, if applicable, the key technical challenges that must be overcome before the technology can be upscaled. Once the experimental, design and operational conditions under which the system performs equally well or better compared to the existing technologies are determined, then the actual scaleup can commence.

*Decontamination of air supply:*

Meeting the challenges of energy efficiency, air flow capacity, and process selectivity requires:

1. *Effective combination of plasma technologies with conventional approaches*, including scrubbing, filtration, etc. allows for a significant decrease of the total process energy cost. In particular, combination of plasma treatment with filtration and scrubbing limits plasma treatment to only charging and partial oxidation of biological and chemical pollutants, thus decreasing electric energy consumption in orders of magnitude.

2. Development of *effective and not-expensive cold plasma power supplies* operating on the power level up to hundreds of kW with an appropriate source that has plasma load matching characteristics.

3. *Plasma-chemistry optimization* to suppress generated non-acceptable treatment byproducts like ozone, nitrogen oxides, etc. It requires, in particular, analysis and characterization of non-uniformity of plasma-chemical processes in air.

To conclude, there is substantial interest in water treatment using electrical discharge plasmas, especially for the degradation of legacy and emerging contaminants or point-of-use fertilizer production. Similarly, the recent COVID-19 pandemic inspired renewed interest in alternative air treatment methods. The *Decadal Assessment of Plasma Science* by the National Academies of Science, Engineering, and Medicine released in May 2020 states: "The central challenge in many of the proposed applications of LTP activated processes to energy, water, food and agriculture is in scaling. Even if plasma can be shown to be effective on small laboratory scales, the process must be scaled sufficiently to make it useful in an industrial setting, whether in the factory or the corporate farm." [22] The feasibility of current plasma devices for the treatment of water and air has been demonstrated on the bench scale; however, their widespread commercial and industrial applicability necessitates further optimization of energy efficiency and the actual scaleup of the devices and their plasma-generating networks to the required levels. For the wide adoption of the plasma-based devices, the process cannot be just as good as the existing technologies but significantly better.

While there have been significant advances in understanding the chemistry and physics of the single and multiphase (gas-liquid) plasma systems, these have been generally studied separately, either from the plasma or from the bulk liquid side. To understand the processes that control chemical conversion and energy efficiency is crucial for the scaleup of any plasma system and requires a holistic multidisciplinary approach that identifies the key measurements, sets of data, and/or the reactor design element(s). In other words, to translate fundamental findings into engineering solutions, the community has to rethink its approach towards the common goal of plasma air and plasma water treatment commercialization.

## V. LTP FOR WOUND HEALING

S. Bekechus

### *V.1 Current State*

Non-healing and chronic wounds are a major burden to both patients and healthcare systems. About two decades ago, the application of low-temperature plasma (LTP) for wound healing was among the first medical problems envisioned to be addressed using gas plasma technologies after finding promising antimicrobial effects of LTP in the 1990s [1-3]. However, taking the concept into the clinical routine took place only in the last decade. The requirements for bench-to-bedside concepts are multifold. The plasma device must be safe, generating a reproducible plasma, manufactured based on quality management procedures, receiving appropriate accreditation by authorities, fulfill reasonable criteria for basing business models on existing needs in the health care market, and, ultimately, medical doctors willing to explore such novel tool in the clinical setting.

The current state of LTP in wound healing can be described from several perspectives, such as i) regional practice, ii) preclinical vs. clinical, iii) and the plasma devices' specifications. From the regional perspective, LTP-mediated wound healing in the clinical setting is most advanced in central Europe, especially Germany. All plasma devices that were accredited by official medical authorities to treat chronic wounds within dermatological centers were engineered and put into practice in Germany. According to the directive on Medical Device Regulation (MDR, EU 2017/745), these devices can be used across the European Union. Official numbers of devices in the field are not available, but based on personal communications, it is estimated that there are a few hundred devices currently sold to or rented by dermatological centers and practitioners. There are efforts in, e.g., the U.S. and the Republic of Korea to achieve similar local authorization for plasma devices aimed at targeting chronic wounds in humans. It is emphasized that this text only focuses on clinical and human cases of wound healing, not on cosmetic or veterinary applications.

For clinical wound healing, several hypotheses exist to explain the benefits from LTP exposure, suggesting several mechanisms simultaneously at play to promote wound healing. Usually being less than 1 log (90%), the antimicrobial efficacy of LTP-treated wounds is documented but overall modest [5,6], also when analyzed ex vivo [7]. Accordingly, a recent prospective, randomized, placebo-controlled, patient-blinded clinical trial comparing plasma vs. antiseptics in a head-to-head study indicated LTP-promoted wound healing independent to antimicrobial efficacy as patients in all groups received antiseptics. The effect of LTP-derived electric fields in wound healing is unclear, especially since there is a lack of studies on intensity-matched exposure of electric fields alone in appropriate in vivo wound models. The mild heat transferred to the wound from the LTP is rather beneficial than detrimental to wound healing [8]. The dominant effector hypothesized to drive LTP-induced wound healing is reactive oxygen and nitrogen species (ROS/RNS) [9,10]. Albeit direct clinical evidence of the roles of ROS/RNS is challenging to obtain, overwhelming data is supporting this idea, as described by D. B. Graves in his 2012 review [11]. In addition to this, recent work in LTP-treated patient wounds suggests plasma to spur on microcirculation and tissue oxygenation in superficial and deeper layers [12,13].

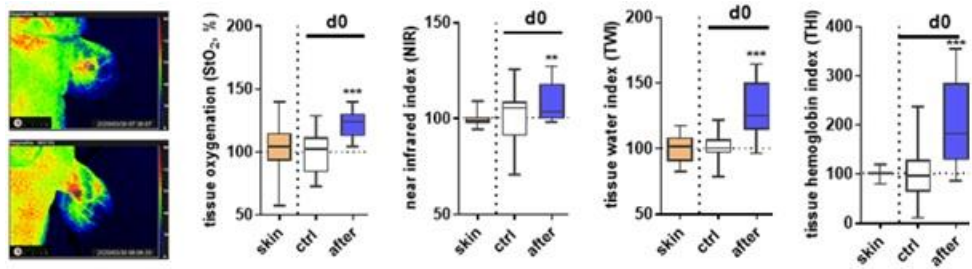

Fig V.1 Hyperspectral imaging of wounded ear (left) with orange wound region in control mice (top) and enhanced signal generation in the plasma-treated wound (bottom). Boxplots show quantitative data for several parameters of hyperspectral imaging. Figure is adapted from [4].

Preclinical evidence of plasma-treated wounds in mice from Schmidt et al. confirmed these findings [14] and suggested an important role of focal adhesions and complexes and matrix remodeling in plasma-treated wounds [4] (see Figure V.1). Most importantly, an array of in vivo data in non-infected wounds of different origins (e.g., punch biopsies, burn, and laser-induced), animal species (e.g., mice, rats, and sheep), and locations (e.g., back and ear) have shown plasma treatment to promote healing in the absence of infection [15-22]. This markedly substantiates the idea of redox control of wound healing via plasma-derived ROS/RNS. The discussion on mechanisms is beyond this article's scope, but it has been recently suggested that the redox-regulating transcription factor Nrf2 and altered immune cell infiltration take part in plasma-accelerated healing using the kINPen [23]. Also, the mechanism of action of the accredited microplaster plasma device was delineated to redox action and stimulation of wound-resident cells of the body [24-27]. Details on in vitro findings were summarized recently [28].

Regarding the perspective on the types of cold plasma sources, only devices in clinical practice are discussed because an overwhelming number of plasma devices have only been covered in relatively few papers and rarely in a clinical setting, while usually lacking accreditation. Of the three devices currently marketed based on thorough clinical experience and medical product guideline compliance or FDA approval specifically for chronic wound healing in dermatology, two are DBDs (PlasmaDerm from Cinogy and SteriPlas from Adtec), and one is a jet (kINPen MED from neoplas med). All have been investigated in clinical trials, as summarized before [29-31]. Two more devices from Germany, PlasmaCare from TerraPlasma (clinical trial completed) and a silicon-based plasma-plaster called CPTpatch from the INP spin-off company ColdPlasmaTech (clinical trial in preparation) are within authorization phases. The DBD device PlasmaCare is innovative because of its battery-driven nature, making it an excellent option for ambulant and home wound care. The CPTpatch is a single-use wound dressing that is planned to be left on the wound to ignite a plasma directly in the wound bed multiple times. This status suggests a dominance of DBD devices over jets in present and future medical plasma wound care. Notwithstanding, clinicians repeatedly emphasize the scalpel-like precision treatment option that comes with plasma jet systems. Especially experienced wound nurses and health care practitioners know that some areas in chronic wounds are more problematic than others [32], which can be effectively targeted with plasma jets in a loco-regionalized manner and, e.g., with individual treatment times. Regardless of the device type, the plasma therapy complements and does not replace existing wound care measures. Critical steps such as wound debridement and dressing change remain to be vital for therapeutic success.

*V.2 Challenges and Proposed Solutions*

The challenges and opportunities in the transdisciplinary field of plasma wound are covering knowledge from physics, biochemistry, and medicine are vast. A detailed discussion, especially individual cell type responses to plasma treatment, is therefore out of scope of this work. The most critical aspects will be discussed nevertheless.

The perhaps greatest asset and caveat at the same time in plasma medicine is the large number of devices reported on. While this is beneficial for exploring the science behind the physics and biology in plasma medicine, it is detrimental to clinical practice. An ideal clinical plasma device would have been accredited across all continents to spur global research efforts, clinical trials, and, ultimately, a certain degree of comparability between results. These would then be independent of the device and depend on treatment parameters, such as treatment duration per cycle, frequency of treatments per week, onset indications of treatment, and offset indications for stopping plasma therapy. Currently, published evidence on such aspects are is absent, hampering the motivation of clinicians less experienced with plasma therapy to implement this technology into their treatment portfolio. Hence, future plasma therapy needs clear guidelines that state the boundary conditions of application, which less plasma-experienced clinicians can also follow. Such guideline is currently set up in Germany within the framework of the so-called AWMF catalog of treatments.. Such medical guidelines would allow reimbursement of plasma therapy, a critical step in the German healthcare system that draws the line between reimbursed standard therapies covered by the obligatory healthcare insurance and non-reimbursed experimental therapies covered by the patient or medical doctor. After all, the additional treatment costs have to significantly reduce the long-term costs of existing wound care to be regarded as superior to existing therapies. In general, low-cost accredited plasma devices commercially available globally will be key for taking plasma wound therapy outside its current niche, with great potential benefits, especially in developing countries.

From a clinician scientist's point of view, several points are unclear in plasma wound therapy, for example: What is the best treatment schedule in terms of duration and frequency? Should the entire wound be treated or, e.g., predominantly infected parts? Are some plasma devices superior in some types of wounds over others? What is the role of infection? Are there common traits in patients that do not respond to plasma wound therapy? Is optimization possible of existing medical product DBD and plasma jet device, and if so, how specifically? What role plays the wound etiology underlying the disease, e.g., diabetes? Should the treatment regimen be adjusted for wound location and size? Is plasma therapy affecting existing wound care measures positively or negatively, and can it be combined with other therapies such as hyperbaric oxygen or vacuum? This shortlist of questions exemplifies the need for applying *and* reporting clinical plasma wound therapy experiences to a much greater extent than currently being the case. Ideally, only a few plasma devices should be used globally in such settings to build knowledge consecutively instead of serially, as is currently the case. Other, although admittedly less complex technologies based on physical principles that had entered medical practice, such as radiotherapy and electrochemotherapy, went similar routes of standardization.

From a preclinical perspective, understanding the interface reactions of plasma with the target tissue is the perhaps least explored but most needed challenge in plasma wound therapy and medicine in general [28]. First of all, it needs to be understood that there is no genuine bulk liquid in wounds. In addition to wound cells and microorganisms, amino acids, proteins, lipids, and nucleic acids will always be close by to the plasma streamers, offering plenty of reaction partners for plasma-derived ROS/RNS. It is therefore plausible that the role of short-lived ROS/RNS is much greater than in liquid-rich laboratory assay systems with cells, in which diffusion distances between the ROS/RNS source (plasma) and the target (cells) are too high for the short-lived ROS/RNS to maintain reactivity. The argument that the tissue is composed of water is also void. A majority of water is in the cytosol of cells, a compartment packed with redox-active proteins, adding to the already complex plasma chemistry. Three conclusions follow from this argumentation. First, it is vital to increase the understanding of which types of ROS/RNS reach the target to what extent. Unfortunately, the number of tools available in redox research to distinguish the different types of ROS/RNS on a quantitative scale is limited. For the fluorescent redox indicators proposed to be specific, virtually all of them have cross-reactivity with multiple types of ROS/RNS, especially when introduced into cells [33,34]. Second, research on non-cellular targets as mentioned above needs to be identified to clarify their role and contribution to plasma-mediated cellular effects. The fact that oxidative modification of such targets can be of biological and immunological relevance has recently been demonstrated for the kINPen [35,36], apart from a broad array of work in the field of redox biology for single ROS/RNS [37-39]. Third, and based on the questions proposed from a clinical perspective, it is critical to understand that only animal models recapitulate the wound environment to a sufficient extent to allow true preclinical conclusions on plasma parameters and biological effectors. The contribution of the immune system, which cannot be accurately mimicked in vitro, is highly critical for the onset and offset of the different wound healing phases to allow complete healing [40-42]. Other vital aspects, such as angiogenesis and matrix remodeling, also take place in appropriate test systems only. By directly comparing infected vs. non-infected wounds, final conclusions may also be drawn towards the role of antimicrobial plasma effects in supporting healing responses. The role of cells of the immune system in plasma-enhanced wound healing might be analyzed using immune-deficient mice or antibody-mediated depletion.

From the plasma device perspective, the field of plasma jets currently offers the greatest opportunity of improvement. Jets have the caveat of being limited to a small area, while patient wounds can be significantly large, leading to long exposure times and ultimately also costs. Several groups have mitigated this issue using multijet arrays [43,44] or are currently working on this. Plasma jets (and gas-driven DBDs) also offer the opportunity of tuning the feed gas towards application-oriented effects, a principle shown multiple times by several groups across in physical, biochemical, antimicrobial, biological, and in vivo characterization [45-50], including wound healing in rodent models [15,51,52]. Ultimately, however, it might be more reasonable to optimize existing plasma devices that have received or are planned to receive accreditation as medical products to facilitate the translation of data and knowledge from physics and biology plasma medicine into clinical plasma medicine.

# VI.  LTP in Dentistry

C. Jiang

*VI.1 Current State*

The earliest application of gas discharges in dental practice dates back to 50 years ago; in 1970s, plasma-sprayed titanium (Ti) or Ti plasma-coating for metals and ceramics were used for oral implants[1]. In the late 80s and early 90s, more applications of low-pressure plasmas in dentistry emerged, including dental instrument sterilization[2-4] and surface treatment of biomaterials such as denture, splint ribbon, composite post, and dentin to assist miscellaneous dental procedures[1, 5, 6]. It was not until the early 2000s, however, direct treatment of teeth and oral surfaces for potential chairside uses became possible due to the development of non-thermal atmospheric pressure plasma (NTAP) based technologies[7]. Applications of plasmas in dentistry have since flourished and entwined with other new technologies in different dental specialties for advancement of the relevant techniques. Reviews or updates on similar topics frequent in dental, medical and engineering journals, especially in the past decade[5, 8-12]. The status/update reported here hence serves as an addition but not a replacement to the previous reviews on the applications of NTAPs in dentistry.

According to the potential clinical use, applications of the plasma-based technology can be categorized as *direct* and *indirect* uses of plasma. Direct uses of plasma refer to applying direct plasma treatment on dentin, gum, implant, or other dental objects of a patient in a dental office for chairside uses. Any other uses of plasma in dental practice such as the use of plasma for cleaning, preparation or modification of dental materials and surfaces, which can be achieved in a laboratory and outside of a dental office, belong to the indirect uses of plasma. Due to the different application settings of direct and indirect uses of plasma, the requirements and challenges for technology adaption are hence significantly different, which also impact the state of the technology development for specific applications.

The direct uses of plasma can be classified as three areas in dentistry: periodontal, endodontic and prosthodontic treatments. While the antimicrobial effects of plasma play the most important roles in the first two areas of applications, surface modification to improve the adhesion properties or change the coloration of dentin is the main cause for the prosthodontic treatment. Table VI.1 lists the specific applications that can be classified in different dental practice areas, the related study models, plasma sources used, and their corresponding status. Periodontal treatments include both removal pathogenic microorganisms or biofilms from cavities (such as caries) or implants that were previously applied in a patient and causing peri-implantitis. For research effort, cavity cleaning may overlap with root canal disinfection, although the choice of microbial strains and treatment substrates may vary. For root canal disinfection, accessibility of the plasma-induced disinfecting effect into the apex and the rest tubular structures of a root canal is an additional challenge for the application, and hence the outcome from *ex vivo* and *in vivo* studies would weigh more to help evaluate the potential success of the technology. The status of both caries removal and endodontic disinfection is hence similar: it is evident that NTAP treatment can be as effective as the use of bleach such as 5-6% NaOCl or 2% CHX, while the combination of NTAP with bleach irrigation can achieve even better disinfection outcome[9, 19-22]. Studies for implant surface cleaning using NTAPs are fewer comparing to other areas of applications. Although the efficacy of atmospheric plasmas, powered by pulsed microwave (MW) or radio frequency (RF), on implant surface decontamination has been demonstrated, safety studies of these plasma applications are lacking. Additionally, a recent study showed that the effect of plasma is not significant for implant surface disinfection when the plasma treatment was added to an air-polishing protocol[16].

Prosthodontic treatments are dental procedures for esthetics or cosmetic restorations. They include tooth bleaching and dentin treatment for bonding to resins. The tooth bleaching application of plasma uses the surface modification properties of plasma and enhances the bleaching activity of bleach agents on dentin surfaces. Studies have shown the combination of NTAP and bleach agents enhanced the bleaching effectiveness by a factor of 3 or higher[26, 27]. For the dentin treatment, a recent systematic review on effect of NTAP on adhesives resin–dentin micro-tensile bond strength (μTBS) showed that NTAP application could enhance resin–dentin μTBS of etch-and-rinse (ER) adhesives or universal adhesives applied in the ER mode[9]. Some studies showed that an optimal dosage (e.g. treatment time) was important to achieve better interfacial bonding strength or μTBS, while overtreatment resulted decreased interfacial bonding strength[30].

The indirect uses of plasma in dentistry include surface modifications of implants for implantology or oral surgery, improving interfacial bonding strength or adhesion of dental objects used in periodontics or prosthodontics, and sterilization of dental instruments. Surface modification using plasmas, both low-pressure and atmospheric pressure, is the most widely used plasma-induced process that impacts various applications including surface modification/activation of Ti, Zr, or CaP-coated implants, dental alloys, improving interfacial bonding strength or adhesion of resins, resin composites, glass fibers, ceramics, polyetherketoneketone (PEKK), and dentures. For decontamination of dental instruments, inactivation of dental pathogens with ion bombardment and oxidation induced by low-pressure non-equilibrium plasmas used to dominate the plasma applications for sterilizing endodontic files or other operative instruments[5, 10]. Although most of these indirect applications started with low-pressure plasmas and may still be able to use low-pressure plasmas, the use of NTAP eliminates the need of a vacuum system, potentially achieves faster cleaning process and can hence reduce the cost of operation substantially. As a result, NTAP has been attracting more and more interests in research and development and may gradually replace the use of low-pressure plasmas in almost every dental field.

The *in vitro* and *in vivo* studies demonstrated that NTAPs could enhance surface colonization and osteoblast activity, and accelerate mineralization of implant (*e.g.*, Ti and Zr) surfaces, as well as improve osseointegration of implant[5, 10, 31]. The antimicrobial effects of NTAPs facilitate biofilm removal and prepare a clean surface of an implant[5, 10]. Additionally, the combination of mechanical and NTAP treatments may result in surface "rejuvenation" against peri-implantitis by activating the synergistic antimicrobial effects and surface improvement[5, 10].

Table VI.1. Direct uses of non-thermal atmospheric pressure plasmas (NTAPs) in dentistry

| Specific Applications | Related study models | Plasma sources (power type, working gas) | Status | Ref. |
|---|---|---|---|---|
| Periodontal Treatments | | | | |
| Removal of dental caries | dental pathogens and pathogenic biofilms grown on well-plates, agar, HA discs, dentins and other substrates, *in vitro* | NTAP plasma jets, pulsed or RF, Ar/He + (O₂) or Air | Large variations in CFU reduction factors depending on the substrates and plasma treatment dosages. In general, plasma effect is comparable with the use of 5-6% NaOCl or 2% CHX irrigation on dentin surfaces. Biofilms are harder to remove. | [8, 13-15] |
| Implant disinfection | single-strain microorganisms (e.g., *S. mitis*, *C. albicans*, *P. gingivalis*) or peri-implantitis biofilms on Ti discs, Ti implants, steel wafers, *in vitro* or *ex vivo* | NTAP plasma jets, MW or RF, Ar/He + (O₂, H₂O) or Air | NTAP was shown effective in removing *P. gingivalis* biofilm from titanium discs, *in vitro*. Comparing with other techniques such as air-polishing, the effect of NTAP is not significant and needs more study. | [16, 17] |
| Endodontic Treatments | | | | |
| Root canal disinfection | *E. faecalis*, *C. albicans* or multi-strain endodontic biofilms grown in root canals or tubular structures mimicking root canals, *in vitro*, *ex vivo*, and *in vivo* | NTAP plasma jets, pulsed or RF, Ar/He + (O₂ or bleach solutions) | In most studies, plasma effect is comparable with the use of ~3% NaOCl or 2% CHX irrigation. Biofilms are harder to remove. | [8, 18-21] |
| Prosthodontic Treatments | | | | |
| Tooth bleaching | Stained or natural extracted human teeth (sectioned or whole), in vitro/ex vivo, rabbit teeth in vivo, were subject to bleach ($H_2O_2$, carbamide peroxide) in solution or gel, combined treatment of plasma and bleach | NTAP plasma jets, low frequency (e.g. 15- 20 kHz) AC or pulsed, He or Ar | Combining NTAP and bleach improved the bleaching efficacy (by a factor of 3 or more) | [22-27] |
| Dentin treatment for ethetics | HA disc or dentin surfaces prepared from extracted human or bovine teeth, after crown removal, followed by etching and rinsing | NTAP glow discharge or DBD, MW, RF, or low frequency AC, He or Ar | Plasma treatment of the dentin surface (for an optimal dosage such as the treatment time) increases the interfacial bonding strength or µTBS, and facilitate the adhesion process | [9, 28-30] |

*VI.2 Challenges and Proposed Solutions*

Decades of research activities have led to substantial development of plasma applications in almost all areas of dentistry. For indirect treatments, the challenges of plasma applications are mainly the efficacy of the plasma-based technologies and the effective cost compared with other non-plasma techniques. The increasing uses of NTAPs are making the plasma-based technology more economically competitive, especially comparing to lasers or UV-light sources required techniques. The highly controlled surface modification by NTAP treatment is becoming the future trend in oral surgery and implantology [10], where chemical functionalization, deposition of antibacterial thin films and coatings, and surface structuring to create antifouling surfaces impact significantly on application outcomes. More importantly, adapting NTAPs into additive manufacturing such as three-dimensional (3D) printing and polymerization to assist multiple areas in dentistry will decisively pin the plasma-based technology in the map of dental technologies. The reports of plasma sterilization of 3D printed dental objects[32] or removal of chemical gradients inside porous structures to enhance cell viability in 3D porous scaffolds (Wan, et al. Biomaterials, 2006) for tissue engineering are promising start for more plasma-processing assisted dentistry [33].

The road to chairside uses of direct plasma treatments, on the other hand, remain more challenging. The large variations in treatment outcome, particularly concerning the antimicrobial effects of NTAPs[8], and the diverse selections of different NTAPs, which can be powered by pulsed DC, RF or MW, present dilemmas for dentistry researchers and students to decide for the next-step clinical trial. Nevertheless, studies designed in a clinically relevant model, such as using multi-species microbial biofilms from patients grown on a dentinal surface or in a root canal as well as animal models have been abundant[13, 18, 19], can be the ultimate testing model for *ex vivo* or *in vivo* studies. Comparisons between NTAP and other non-plasma treatment protocols are important. Comparable antimicrobial effects of NTAPs against dental biofilms in *ex vivo* root canals or animal models are great milestones in plasma applications for periodontal or endodontic treatments. Close collaborations between the dentists and plasma engineers are essential to identify the effective treatment protocols and efficiently use resources for technology advancement. The other challenge is the safety concern. The reactive oxygen and nitrogen species (RONS) produced by NTAPs have been considered playing important roles in antimicrobial processes and surface modification, and also known to cause oxidative and nitrosative stress to cells and tissue. Nam *et al.* recently conducted a histological safety study in a rabbit model using a NTAP for tooth bleaching [34]. The NTAP was based on a dielectric barrier discharge configuration, powered by 15 kHz AC at 10-kV peak voltage. The working gas was Ar. The gap temperature was about 35°C. The study concluded that the combinational treatment of plasma and 15% carbamide peroxide (CP) resulted in the most color change compared with 15% CP treatment alone, and the combination treatment did not indicate inflammatory responses nor thermal damages[34]. The authors also noted that the plasma did not cause histological damage in oral soft tissues during tooth bleaching[34]. Another recent *in vivo* study conducted by Yao *et al.* demonstrated the antimicrobial effectiveness of a NTAP against apical periodontitis in a beagle model, comparable to that of 2% CHX irrigation, with no detected structural damage in dentin or tissue necrosis at the periapical region[18]. The NTAP was generated from a handheld plasma jet device, using He as the primary working gas, and powered by pulsed DC at 10 kV and 10 kHz[18]. The authors commented that modifying the NTAP jet by following He through 2% CHX enhanced the antimicrobial efficacy[18]. However, the safety study on the modified method was not reported. After all, the generation of RONS were associated with many technologies in dentistry including lasers, photosensitizers, bleaching agents, and even resin cements [35]. Hence NTAP is not the only technology may cause safety concerns. The critical point here is the dosage control. More safety studies similar to the above two examples would help pave the way carrying the NTAP-based technology through clinical trial and towards the dental chair.

# VII.  LTP FOR CANCER TREATMENT

M. Keidar, V. Miller, S. Bekeschus, D. Yan, H. Tanaka, M. Laroussi, K. Ostrikov, and M. Hori

### *VII.1 Current State*

In recent years low-temperature plasma, LTP, is increasingly being investigated as a possible new modality for cancer treatment [1], [2]. The novelty of LTP lies in its multi-factorial effects that include reactive oxygen and nitrogen species (ROS/RNS) produced in the plasma, physical factors like emitted electromagnetic waves and the electric fields that are formed when plasma impinges on tissue [3]. Rapid advances in LTP technology have allowed great control on the final composition and properties of plasma for specific biological processes aimed at clinical applications. Specifically in oncology, LTP has enabled scientists to explore the mechanisms of eliminating cancer cells that would be applicable in diverse types of cancers. Focal solid tumors require different treatment approaches than more diffuse or metastatic tumors. Broadly, the experimental approaches encompass two methods of inducing cancer cell death: direct killing of cancer cells by LTP *in situ* and stimulation of immune responses by inducing controlled oxidative stress in cancer cells. Numerous studies use *in vitro* 2-D cell culture models, and the use of 3-D, *in-ovo,* and animal models that are very much in alignment with the cutting-edge approaches in oncology therapeutics, is increasingly being reported. These studies expand depth and breadth of knowledge about the efficacy of LTP in producing the desired anti-cancer effects and the mechanisms of LTP action further enabling improved methods of plasma delivery. The encouraging aspect of these investigations is that an initial proof-of-concept clinical case study has already been performed [4], and its results can be utilized to design studies that ask practical questions [5].

The widely accepted hypothesis of LTP interaction with cells and tissue is based on the notion that chemical elements of LTP are stimulating or toxic, depending on their concentration [6]. Various unique combinations of these species might provide great potential for activation of specific signaling pathways in cells where no single component is yet identified as the key effector. LTP application to tumor cells can have cytotoxic effects both *in vitro* and *in vivo*. Some studies suggest that non-malignant cells are less sensitive to the same regime of LTP in some studies, though this is not necessarily always the case [7]–[9]. These outcomes are believed to be related to the chemical species in plasma. The sensitivity of cancer cells to either direct treatment or indirect treatment is proposed to be due to redox imbalance and to specific antioxidant systems that are linked to the expression of tumor suppressor genes like p53 [10].

The basic strategy of indirect LTP treatment is the application of plasma-treated liquids (PTS) such as cell culture medium or aqueous solutions like lactate to kill cancer cells or destroy tumor tissues. The anticancer effects are achieved by the chemical components, including reactive species and other chemically modified substances in these solutions [11]–[13]. Typically, PTS are produced through direct contact of LTP with the solution during treatment. However, PTS can also be produced by the gas discharge within the solution [14], [15]. PTS has been shown to selectively kill many cancer cell lines *in vitro* or have anti-tumor effects when directly injected in subcutaneous or intraperitoneal tumor models *in vivo* [16]–[19]. Once produced, PTS is independent of plasma source, which is an advantage over direct plasma treatment. Treatment by PTS eliminates the local effects of electric fields, photons, and temperature. Some cancer cell lines, such as the melanoma cell line B16F10, show only modest cytotoxicity to the reactive species in PTS [20]. PTS also shows promising application in the treatment of metastatic cancers as shown in a recent study that addresses epithelial-mesenchymal transition (EMT). EMT is a process which endows cancer cells with increased stemness, metastatic potential, and resistance to conventional therapies. In this study PTS was shown to be significantly more effective against cancer cells that have undergone EMT than their epithelial parent cells. This effect is hypothesized to be due to the increased ROS level in the mesenchymally-transitioned cells [21]. As a potential pharmacological modality, the stability of PTS is an important consideration needing further investigation [22]–[24].

Unlike indirect LTP treatment, direct plasma treatment is performed when cells or samples are directly accessible to bulk plasma [25]. A layer of physiological solution typically covers the cells during direct *in vitro* LTP treatments [26]. Such a layer facilitates the diffusion and formation of reactive species from the gas phase to the liquid phase in the medium to produce desired biological changes in cells [27]. Physical factors in LTP, such as thermal effect, UV, and electric fields, may be largely blocked by such an aqueous layer, particularly when it is adequately deep. Direct and indirect treatment share many similarities [28]: many cellular responses have been widely observed in both treatment strategies, including perturbations in the cell membrane, changes in mitochondrial structure and function (increase in intracellular ROS), and damage to the endoplasmic reticulum and DNA [29]. Cell death is the final goal and fate of most LTP-treated cancer cells [30], [31]. Apoptosis is the primary form of cell death; however, necrosis and autophagy-associated cell death have been also reported [32]–[35].

Direct treatment takes advantage of two unique components: short-lived reactive species and physical factors (such as electromagnetic emission from plasma) otherwise absent in PTS. Under the same experimental conditions, the cytotoxicity of direct treatment is sometimes stronger than indirect treatment [36], [37]. Direct LTP treatment is also suggested to produce additional cellular effects like endogenous $H_2O_2$ generation by cells [38]–[41] and sensitization of cancer cells to the cytotoxicity of long-lived reactive plasma species such as $H_2O_2$ and $NO_2^-$ [42], contributing to the uniqueness of LTP. Physical plasma factors alone can also sensitize cancer cells to the cytotoxic action of ROS, as seen for nanosecond-pulsed magnetic fields against B16F10 melanoma cells [41] and pulsed electric fields against leukemia cells [42]. A recent study demonstrated that glioblastoma cell lines U87MG and A172 could be sensitized to cytotoxicity of temozolomide (TMZ), a drug used for treatment of brain tumors, by the electromagnetic emission from a helium-discharged tube [43]. Vice versa, drug-sensitized melanoma cells have been shown to be more susceptible to plasma-induced cell death [44].

Direct LTP treatment has already been widely used in many animal studies. Direct plasma treatment on the skin above a subcutaneous glioblastoma U87MG xenograft tumor site in mice reduced tumor size and prolonged their life [45]. Fractionated treatment was demonstrated to have improved anti-tumor effect compared to a single, longer treatment [46]. Similar anti-tumor effects after direct plasma treatment of subcutaneous tumors of different origins, including melanoma, bladder cancer, pancreatic carcinoma, breast cancer, head and neck cancer, and neuroblastoma in animal models have been reported [47]–[50].

VII.2 Challenges and Proposed Solutions

A key challenge in the use of LTP for treatment of cancers is the identification of its mechanism of action. The contribution of individual plasma components to its antitumor effects remains largely unknown. While most of the research describes the cause-and-effect relationship between chemical species and tumor toxicity, it is still unclear how much the physical factors of LTP contribute to the cellular responses observed. An important, practical consideration to keep in mind is that when cancer cells or tissues are directly exposed to LTP, they experience both physical and chemical effectors simultaneously and as such, the observed biological response could be attributed to the synergistic effect of the physical and chemical factors. *In vitro*, these effects are difficult to study because liquid layers of dozen of micrometers block physical factors and only allow chemical factors such as reactive species to affect cells. An initial attempt was recently published where cells were treated from the bottom of an inverted multi-well plate or cell culture dish to block all factors except electric fields [51]. This *in vitro* treatment caused necrotic killing effect on at least six cancer cell lines, including cell lines that are resistant to reactive species [52], [53]. *In vivo*, necrosis may trigger an immune response or an inflammatory response [54]. where physical factors in LTP are very relevant in direct treatment strategies. Furthermore, based on the results of a recent study, the effects of microwave emission from a plasma was determined to have decontamination efficacy [55]. This trans-barrier capacity of LTP treatment suggests the non-invasive potential in medical application but more extensive studies are needed to understand the mechanism of interaction of physically-based modalities with cells and tissues.

The second challenge is the insufficient understanding of the biological impact of LTP on tissues, a prerequisite to explore the underlying mechanism of LTP-based anti-cancer effects. Besides the well-accepted direct cytotoxic effects of LTP on cancer cells, evidence is now accumulating to show that it triggers unique immune responses and immunogenic cell death *in vivo*, providing clues to help understand LTP effects at the tissue level. Immunogenic cancer cell death (ICD) is elicited in response to various stimuli [56]. Sterile inflammation via damage-associated molecular patterns (DAMPs) overrides the immunosuppressive properties associated with apoptosis, which drives dendritic cell (DC) maturation and anti-tumor T cell responses [57]. The importance of immunity [58], immunogenicity and antigen presentation [58], ICD [59], and vaccination [60], [61] for plasma cancer treatment has been extensively reviewed recently. Increased immunogenicity has been described in several *in vitro* studies [62]–[69], and a number of reports suggest an enhancement of anti-tumor immune function following plasma cancer treatment *in vivo* [70]–[73]. The mechanism proposed is that plasma-induced oxidative stress elicits ICD, albeit direct *in vivo* evidence, e.g., via systemically enhanced antioxidants or tumor cell clonotypes with enhanced expression of antioxidant enzymes to scavenge ROS/RNS and thus inhibit (immunogenic) cell death, is lacking. From recent *in vitro* data, it can be concluded that the role of antioxidant defense molecules such as catalase is less pronounced in terms of plasma-induced cell death induction [74]. *In vitro*, several studies have suggested activation of professional antigen-presenting cells (APCs), myeloid cells critical in providing T cell co-stimulation and activation, following plasma treatment or following exposure to plasma-treated tumor cells [75]–[80]. A new research field and potential therapeutic approach has been recently opened up by investigating the immunogenicity of proteins, showing that plasma-treated chicken ovalbumin, a model protein, provided enhanced protein immuno-recognition *in vitro* and tumor growth reduction *in vivo* [75]. The implications of these findings for tumor vaccination and therapeutics are vast and provide a basis for further investigation [61].

The ultimate goal of plasma oncotherapy is its translation for clinical application and it remains an important topic for continued exploration and discussion among experts in the multiple disciplines that contribute to exciting new developments in plasma medicine. Based on the different ways by which tumor tissue may be exposed to LTP, as decribed above, it is quite clear that it offers great flexibility in different cancer scenarios. Here we propose our vision for how LTP may be used for cancer therapy.

For the more accessible tumors, like those that are subcutaneous, a direct, fractionated treatment with few to several treatment cycles over the skin of the tumor site would be easy and practical. Preliminary reports using the kINPen MED plasma jet in patients suffering from metastatic melanoma was however unsuccessful, perhaps because of the thickness of the skin. In contrast, treatment of ulcerating, locally advanced squamous cell carcinoma of the oropharynx [4] inhibited the growth of tumors in some patients. Incisional biopsies found apoptotic tumor cells and a desmoplastic reaction of connective tissue. In another clinical application, LTP was used to treat oral premalignant lesions (leukoplakia) [81]. Side effects were mild, not life-threatening, and transient with no long-term after effects [82]. Improved understanding of the mechanism of LTP action could help improve clinical treatment guidelines for better efficacy. In the case of deep tumors like intraperitoneal carcinomatosis, while topical LTP treatment is not possible, PTL offers a convenient solution via subcutaneous or intraperitoneal injections. The sensitization of cancer cells to the cytotoxicity of drugs provides the added benefit of a third option of synergy with LTP, as previously suggested in *in vitro* and translational models using the guideline-based drugs gemcitabine and cisplatin [83]. This LTP-induced selective sensitization of cancer cells is another new direction for further investigations: not aim to kill cancer cells but to sensitize cancer cells to traditional chemotherapy by chemical or physical factors as well as their combinations.

LTP may be also be used as an adjunct therapy to standard of care surgery by treating exposed tumor tissues or the remnants of tumor tissue during surgery. In a minimally invasive surgical approach, LTP can be used with endoscopic plasma technology to treat deep tumor tissues [84].

LTP as an adjunct therapy was tested clinically for the first time in a patient with stage 4 colon cancer at Baton Rouge General Medical Center in Baton Rouge, Louisiana, immediately after surgery to remove the tumor [1]. Phase I safety clinical trial was completed in 2021 that involved 20 patients to evaluate the safety of the procedure [85].

An important consideration for translation is the precise and controlled delivery of plasma effectors. The chemical composition, physical emission, and other parameters of LTP can be quickly modulated for specific needs [86]–[90]. To this end, an intelligent LTP system should scan the cellular responses to LTP and modify discharge conditions in real-time via feedback mechanism based on machine learning [91]. A real-time control of LTP is capable of optimizing the killing effect on cancer cells or tissues while protecting normal cells or tissues [92]. Such an intelligent LTP system is proposed as the self-adaptive LTP source, which may guide the design of the next generation of LTP sources [93]. To build an adaptive LTP system, a multi-parametric control system is necessary. The feedback control algorithm may use nonlinear model-predictive controls (MPC), a strategy used to implement a numerical solution of an open-loop optimization in feedback [94], [95]. The adaptive plasma approach may ultimately realize a personalized LTP-based cancer therapeutic that could be adapted for treatment of other diseases.

## VIII. MODELING OF LTP-CELL AND LTP-TISSUE INTERACTIONS

A. Bogaerts[*] and M. Yusupov

*VIII.1 Current State*

For improving plasma medicine applications, a good insight is needed in the interactions of LTP with cells and tissues. This can be obtained by modeling, as it can provide detailed information, even on the molecular level, which might experimentally be less (or not) accessible. Below, we make a distinction between modeling the interaction of LTP with (i) entire cells or tissue, and (ii) individual cell components, because they require quite different modeling approaches.

*Modeling the interaction of LTP with entire cells or tissue:*

The interaction of LTP with entire cells or tissue can up to now only be simulated on a macroscopic scale. As a result, the information is less detailed, but it can give an approximate view on the effect of e.g., electric fields induced by plasma streamers on tissue, as demonstrated by Babaeva and Kushner [1] for a DBD plasma interacting with wounded skin. The actual skin tissue was mimicked by a cellular structure in the model. In collaboration with Ning and Graves, the authors also calculated the ion energy and angular distributions incident on cell membranes, and applied molecular dynamics (MD) simulations of ion sputter probabilities of typical lipid-like material [2]. In some follow-up papers (e.g., [3, 4]), Babaeva, Tian, Lietz and Kushner extended their model to wounded skin covered by liquid, containing also blood platelets. The major focus was on the plasma-liquid model, but some description was also given on the electric fields delivered to the blood platelets and cells [3], and on the interaction of the DBD treating liquid-covered tissue [4]. Likewise, Kong and coworkers [5] developed a semi-1D model for plasma-biofilm interaction and plasma-tissue interaction, based on a reactive penetration model for mass transfer of plasma species across the gas-liquid boundary.

*Modeling the interaction of LTP with individual cell components:*

Modeling LTP interactions with individual cell components is possible on the molecular level, and much progress has been made in recent years. Depending on the system size and level of detail, different modeling approaches can be used, ranging from quantum mechanics (QM), over density functional theory (DFT) and density functional tight binding (DFTB), to classical molecular dynamics (MD); see [6] for a detailed description. DFT (and other QM methods with higher accuracy) are typically too time-consuming for describing realistic model systems of interest for plasma medicine, but DFTB has been applied, e.g., to study the interaction of ROS with the head group of the PLB [7]. Among the classical MD methods, we distinguish reactive and non-reactive MD simulations. Reactive MD allows the breaking and formation of bonds, and thus the description of chemical reactions between plasma species and biomolecules, as illustrated, e.g., for the interaction of ROS with peptidoglycan [8] (of interest for bacteria killing), with lipids [2] and DNA [9]. Non-reactive MD, also called "molecular mechanics" (MM), does not consider bond breaking and formation, but it allows to describe large systems and over longer time scales, e.g., for the transport of species through a PLB or the structural deformation and stability of biomolecules. This method has become increasingly attractive in plasma medicine, and some examples will be presented in next sections.

*(a) Modeling LTP interaction with the phospholipid bilayer (PLB):*

When RONS produced by LTP come in contact with cells, they first "see" the cell membrane, composed of a PLB with proteins embedded. It is of great interest to study the permeability of RONS through the PLB. It can provide information on RONS distribution, mobility and residence times at the membrane-water interface, as well as their translocation free energy barriers across the PLB, showing whether they can easily reach the cell interior. For instance, Cordeiro [10] studied the transport properties of ROS (i.e., OH, $HO_2$, $H_2O_2$ and $O_2$) across the PLB, employing non-reactive MD simulations. He found that hydrophobic $O_2$ is able to accumulate at the PLB interior, whereas hydrophilic OH, $HO_2$ and $H_2O_2$ mainly stay around the lipid head groups [10]. Furthermore, the PLB can also be oxidized by LTP, and therefore, simulations have been performed to compare the RONS permeability through both native and oxidized PLBs [11]. Figure VIII.1 illustrates the free energy profiles (FEPs), obtained from so-called umbrella sampling (US) non-reactive MD simulations, of various RONS across both native and 50% oxidized PLBs. The FEP of hydrophilic ROS (i.e., OH, $HO_2$, $H_2O_2$) shows a minimum around the head groups, followed by a steep rise towards the center, thus creating a clear energy barrier to cross the PLB. These results are similar to those mentioned above [10]. Upon oxidation, the hydrophilicity of the PLB rises, resulting in a larger permeability for hydrophilic ROS, as demonstrated by the much lower free energy barrier in Figure VIII.1(b).

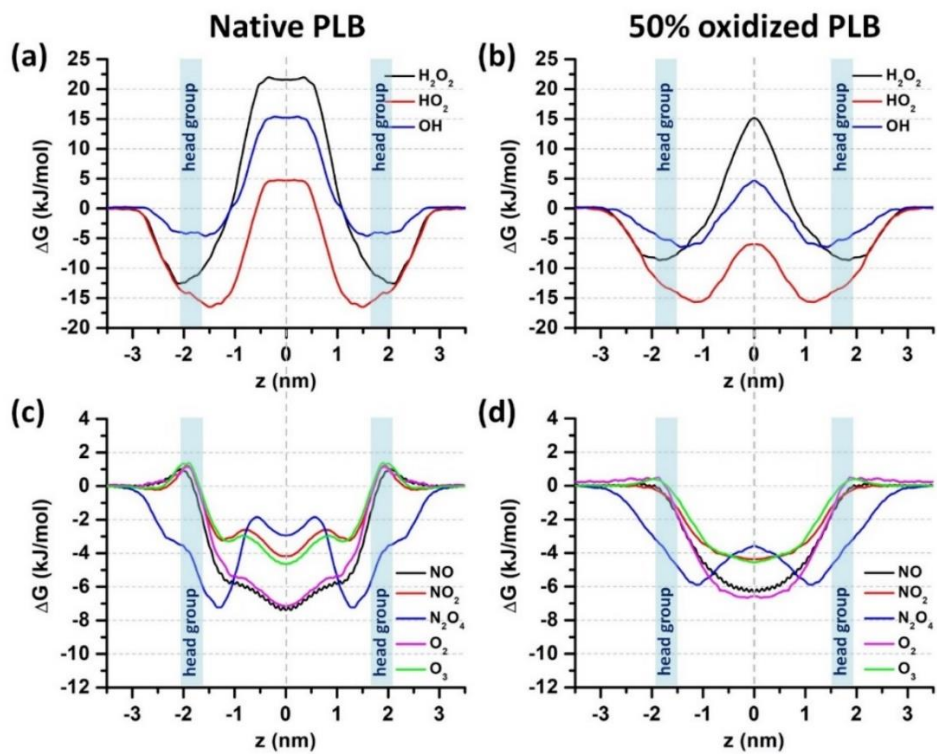

Fig VIII.1 Free energy profiles (FEPs) of the hydrophilic (a-b) and hydrophobic (c-d) RONS, across native (a,c) and 50% oxidized (b,d) PLBs. The average positions of the head group regions (i.e. phosphate atoms of the PLB) are indicated by light blue color.

Because the hydrophilic ROS prefer to reside close to the head groups, DFTB simulations were performed to study oxidation of these head groups [7]. $HO_2$ and $H_2O_2$ were found not to chemically react with the head groups, while OH radicals cause detachment of some parts in the lipids, enhancing the membrane fluidity and reducing the lipid order. This may allow RONS to penetrate more easily through the PLB, causing further lipid tail (per)oxidation, which might result in pore formation (see further).

The hydrophobic RONS (e.g., $O_2$, $O_3$, NO, $NO_2$ and $N_2O_4$) exhibit very low permeation barriers around the PLB head groups and minima in the center; see Figure 1(c,d). Hence, they can cause lipid (per)oxidation in the lipid tail region. Upon oxidation, the PLB can become a bit more fluidic, making the FEPs of the hydrophobic RONS a bit smoother, but otherwise they do not change significantly. These simulations correlate well with experiments, where the permeability of hydrophobic RONS (NO and $O_2$) was found to be 3-6 orders of magnitude higher than for $H_2O_2$ [12]. Thus, hydrophobic RONS may easily permeate through the cell membrane, while the active transport of hydrophilic ROS in and out of the cell should only be possible in the presence of aquaporin (AQP) channels or pores. US simulations [13] indeed predicted a 100 times higher permeability through AQP than through the PLB, which might explain the selectivity of LTP towards cancer cells, because cancer cells typically have a higher AQP expression in their cell membrane. In addition, pore formation in the cell membrane after lipid (per)oxidation has also been studied by non-reactive MD [14], as well as the combined effect of lipid oxidation and an electric field [15].

Besides the higher expression of AQP, some cancer cells also have a lower cholesterol content in their cell membrane than normal cells. MD simulations [16] revealed higher free energy barriers for hydrophilic ROS at higher cholesterol contents, making it more difficult for them to penetrate through the cell membrane. Furthermore, the FEP of $O_2$ showed new free energy barriers close to the sterol rings, which will limit the probability of lipid (per)oxidation of the lipid tails, and hence pore formation. Finally, MD simulations also predicted the absence of pore formation for cholesterol fractions above 15 % [14]. Hence, RONS should be able to penetrate more easily through the membrane of cells with lower cholesterol fraction, such as cancer cells, resulting in oxidative stress inside the cell, while this effect might not happen in normal cells, due to their higher cholesterol fraction. This might provide another explanation for the selective anti-cancer effect of LTP.

*(b) Modeling LTP interaction with DNA :*

Reactive MD simulations revealed two main types of reactions of OH radicals with a DNA model system, i.e., H-abstraction and (ii) OH-addition [9]. The H-abstractions create a radical, resulting in even more (intermolecular) H-abstractions, possibly leading to single strand breaks (SSBs). Combined with a second SSB at the opposite strand in close vicinity, this may lead to a double strand break (DSB). The latter could however not be observed in these MD simulations, because of their limited time scale (see also section 2). The OH-addition on the purine ring of the nucleic bases, more specifically at the C-8' position of dAMP and dGMP, yielded 8-hydroxy-purine adduct radicals (8-OH-Ade$^\bullet$ or 8-OH-Gua$^\bullet$), which is the first step towards 8-oxo-guanine (8-O-Gua) and 2,6-diamino-4-hydroxy-5-formamidopyrimidine (FapydG) formation, i.e., markers for oxidative stress in cells. The latter may affect biochemical pathways within the affected cell, e.g., introduction of DNA mutations or inhibition of gene expressions, possibly leading to apoptosis, but again, the simulated time was not long enough to observe these further reactions, due to the high computational cost of the simulations.

*(c) Modeling LTP interaction with proteins :*

Non-reactive MD, molecular docking and US (binding free energy) calculations investigated the structural conformation and binding affinity of human epidermal growth factor (hEGF; one of the important signaling proteins in both wound healing and cancer treatment) to its receptor (hEGFR) under oxidative stress as induced by LTP [17]. Mild oxidation was found to not significantly affect this binding affinity, and will thus not strongly influence cell signaling pathways, and hence cell proliferation, which might explain why short plasma treatment times are beneficial for chronic wound healing. On the other hand, the interaction was clearly reduced upon stronger oxidation, which might disturb the cell signaling pathways, ultimately leading to disruption of cell proliferation. This may explain why longer plasma treatment times result in inhibiting cancer cell proliferation and even cancer cell death [17].

In general, the effects of post-translational modifications (e.g., methylation, phosphorylation, glycosylation and acetylation) on various protein structures have been widely studied in literature, but it would also be interesting to study in more detail the effects of oxidation by ROS (and nitrosylation by RNS), on the protein stability and its binding affinity, specifically in the context of LTP for cancer treatment, which is till now only scarcely investigated. Indeed, such studies would help to further unravel the molecular level mechanisms of cancer cell death induced upon oxidation by LTP.

## VIII.2 Challenges and Proposed Solutions

*Compromise between level of detail and calculation time:*

The major challenge in modeling LTP interactions with cells or tissues is the compromise between calculation time and level of detail of the simulations. Macroscopic modeling is fast, but can only provide a general, approximate picture. Molecular-level modeling yields detailed information, but the small time step needed makes these simulations limited to small system sizes and time scales. In addition, the complexity of cells or tissues makes it difficult to study the processes at the molecular level. Thus, important and specific parts of cells (e.g., the PLB or transmembrane proteins) are often chosen as model systems in molecular-level simulations in order to understand the underlying mechanisms of the LTP effect on cancer cells. Figure VIII.2 presents the time and length scales of the various molecular-level simulations, mentioned in the previous section.

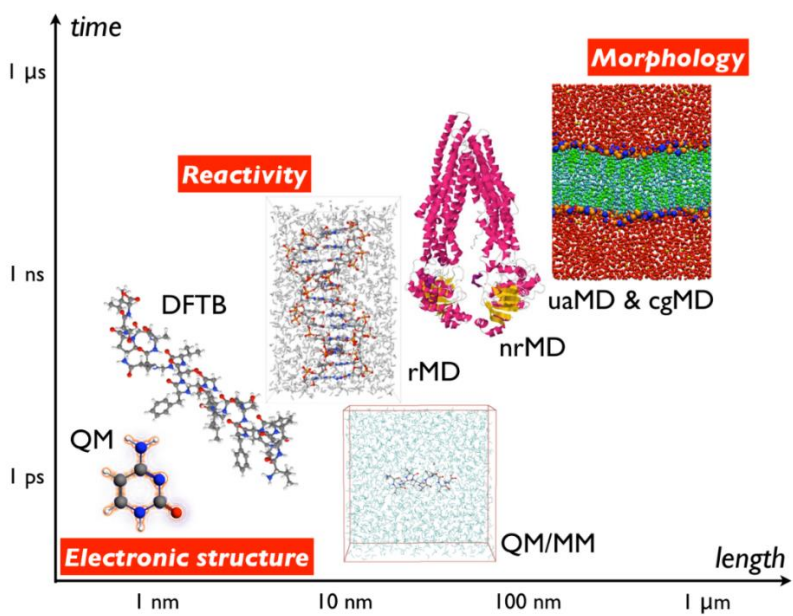

Fig. VIII.2 Overview of the various molecular-level modeling approaches for the interaction of reactive species (e.g., RONS) as produced by LTP, with biomolecules, or for the (longer-term) effects of these interactions, illustrating the length and time scales that can be reached. QM = quantum mechanics, DFTB = density functional-tight binding, QM/MM = quantum mechanics / molecular mechanics, rMD = reactive molecular dynamics, nrMD = non-reactive molecular dynamics, uaMD = united-atom molecular dynamics, cgMD = coarse-grained molecular dynamics. The length and time ranges are approximate and improve over time with the increase of supercomputer capacities. Adopted from [6] with permission.

The computational quantum mechanics (QM) methods, including DFT, are the most accurate, but for plasma medicine, only approximate DFT methods (e.g., DFTB) or classical reactive MD methods would be feasible, keeping in mind the required system sizes of relevant biomolecules. While DFT is more approximate than other QM methods, it is also still very time-consuming, and thus able to handle system sizes only in the order of 100 atoms for a simulation time ranging from a few days up to a few weeks (on e.g., a supercomputer node with dual 14-core Intel E5-2680v4 Broadwell generation CPUs connected through an EDR InfiniBand network). Note that the time range depends on which type of software (e.g., VASP or Gaussian) is employed and whether a simple geometry optimization or deeper characterization is carried out, as well as on the system size, i.e., finite (e.g., cluster) or infinite (i.e., system in a periodic box). DFT-based MD calculations, also called "ab initio MD" (AIMD) can handle time scales in the order of picoseconds. DFTB, on the other hand, can typically handle a few 1000 atoms on time scales of 10s of picoseconds, making it indeed more suitable for biomedical studies on relevant model systems. Classical reactive MD, in which the forces are not based on QM data but on classical fitting parameters, can handle much longer time scales (order of 1 ps to 100 ns) and larger systems ($10^3$-$10^5$ atoms), depending on the complexity of the force field (interaction potential), but of course, at the expense of the level of detail. Moreover, the accuracy critically depends on the chosen force fields. Because of the larger system sizes, it has been used more often in literature in the context of plasma medicine (see section 1.2). However, the simulation time scale is still very limited, and thus, many relevant biochemical processes are still beyond reach, as illustrated in section 1.2(b), for the interaction of LTP with DNA. Indeed, DSBs in DNA, and the reaction of 8-OH-Ade$^{\bullet}$ and 8-OH-Gua into 8-O-Gua and FapydG (i.e., markers for oxidative stress in cells), could not be observed within the time scale of the MD simulations.

Non-reactive MD can handle much larger system sizes and time scales (typically two orders of magnitude) than reactive MD. So-called "all-atom force fields" (i.e., when all atoms in the system are treated separately), can describe in the order of $10^5$-$10^7$ atoms, at time scales of 0.1 ns – 10 µs. So-called "united-atom force fields" treat all heavy atoms separately, while the H atoms bound to a C atom (e.g. in the apolar tails of phospholipids) are treated as one (methyl or methylene) group. Hence, this yields less separate particles in the system, allowing to simulate larger systems (typically up to one order of magnitude larger), for the same time scales. Finally, "coarse-grained methods", in which the atoms comprising entire functional groups (i.e., typically, 3-5 heavy atoms with their H atoms) are represented by so-called coarse-grained particles, further reduce the number of particles in the system, and thus allow even larger system sizes (or to speed up the calculations). This opens possibilities for describing larger biomedical systems, albeit at the expense of the level of detail.

In general, there will always be a trade-off between calculation time and level of detail/accuracy. Thus, it can be interesting to combine the above methods, for a more comprehensive picture. For instance, Yusupov *et al.* [7] combined reactive (DFTB) MD (to study the interaction of ROS with the head groups of the PLB) with non-reactive (classical) MD (for the subsequent effect of head group and lipid tail oxidation on the structural and dynamic properties of the cell membrane).

Furthermore, while the above example illustrates the use of two separate methods, it is also possible to combine different methods in an integrated simulation, in quantum mechanical / molecular mechanics (QM/MM) methods. A small (chemically most relevant) part of the system (e.g., the active site of the biological system) can be described at the quantum chemical (electronic) level, while the surrounding embedding atoms and molecules may be treated at a classical (atomic) level. This method can thus combine the strengths of different modeling approaches, and it would be very interesting to explore its potential in plasma medicine. Indeed, QM/MM is successfully used for many years in biochemistry, e.g., to describe the enzymatic reaction mechanisms (or catalytic properties) of various enzymes [18].

Finally, it would be interesting to apply other enhanced sampling methods in the context of plasma medicine. These methods include US (see section 1.2(a)), metadynamics (MetaD), steered MD (SMD), replica exchange MD (REMD), and accelerated MD (aMD), which are successfully applied to a wide range of biological systems, to study conformational changes of proteins, protein-protein and protein-ligand interactions, enzyme catalysis, and so on [19]. Each of these approaches has its advantages and disadvantages from the computational and application perspective. For instance, US is very popular for free energy barrier calculations, because of its quick convergence and ability of simultaneously running multiple independent simulations (windows). However, the harmonic potentials used for US windows need to be tuned manually, to obtain reasonable positional overlaps between the windows along the collective variable (CV, e.g., bilayer normal). Moreover, adding further US windows to the system may be necessary to improve the convergence of the sampling. Lastly, US simulations are time consuming for large and complex systems. An advantage of MetaD is that *a priori* knowledge of the end states is not required, and several CVs can be used to explore possible pathways. However, the efficiency of MetaD does not scale well with the number of used CVs, and the accuracy and convergence depends on the choice of the necessary parameters [19]. When little is known about the events to be simulated (e.g., protein folding), REMD or aMD are probably the best methods. More information on these and other enhanced sampling and free energy calculation methods is given in [19].

*The need for accurate input data:*

Another challenge, related to the first one, is the need for accurate input data in both macroscopic and molecular-level simulations. This certainly applies to the macroscopic simulations, which are based on many assumptions, but even for the most accurate molecular-level methods, the accuracy of the calculation results critically depends on the input data, e.g., the chosen functionals in DFT or the force fields in classical MD. Hence, there is a constant need for improved functionals and force fields, to further improve the predictive power of the calculations. Even more, the lack of some force fields is limiting the possibilities of MD to obtain an overall picture. For instance, the lack of accurate force field parameters of RNS (e.g., $HNO_2$, $HNO_3$, $ONOOH$) for reactive MD simulations makes it up to now impossible to study their interaction with cell components.

*Experimental validation:*

Experimental validation is another big challenge, certainly for the molecular-level simulations. Indeed, due to small sizes of biomolecular systems, sophisticated experimental methods, such as neutron scattering, small-angle X-ray scattering (SAXS), nuclear magnetic resonance (NMR), cryogenic electron microscopy (cryo-EM), and electrospray ionization mass spectrometry (ESI-MS), would be needed. Furthermore, dedicated experiments should be designed, under very controlled conditions, generating for instance only a beam of OH radicals, instead of the cocktail of RONS produced by LTP. Moreover, experiments should be performed for well-defined model systems, as considered in the simulations, gradually mimicking more complex tissues. Such experiments are set up in various labs, for instance, for studying the separate and synergistic effects of plasma-generated radicals and UV/VUV photons at the cellular and molecular level for various kinds of biomolecules, or experiments with simple model systems for the cell membrane, based on synthetic phospholipid membrane vesicles or liposomal model membranes (e.g., [20-22]).

For instance, in [7], the MD simulations were validated experimentally, using liposomes as model systems for the biological membranes. The so-called Laurdan Assay was used to determine the phase states of the membranes (gel- or liquid-phase) by measuring the penetration of water molecules into the bilayer, which strongly correlates with the packing of the phospholipids. In addition, high resolution mass spectrometry was applied for studying modifications of the liposomes after LTP treatment. The experiments revealed a slight initial rise in membrane rigidity, followed by a strong and persistent increase in fluidity (indicating a drop in lipid order) upon LTP treatment, in good agreement with the simulations. Besides, also indirect validation, by means of biological assays, can be very useful, and a few examples are presented in the next section.

*Translating the insights for improving biomedical applications:*

Finally, while the molecular-level simulations provide detailed insights in e.g., how RONS can penetrate through the cell membrane, either by passive transport (for hydrophobic RONS) or through pores or AQP channels (for hydrophilic RONS), the long computation time dictates the use of simple model systems (e.g., a simple PLB and/or AQP), while the cell membrane contains many other lipids and proteins as well, such as antiporters, catalase, etc. Indeed, the reality in plasma medicine is much more complicated, and it is not straightforward to translate the insights obtained from these simple model systems to the real world, and thus to actually improve the biomedical applications. For instance, taking the same example of the cell membrane, the various RONS might create a myriad of different lipid oxidation (and nitration) products, which are not yet accounted for in these simulations. Hence, more knowledge is crucial on the different products formed, and how they affect the biophysical properties and function of the cell membrane. Furthermore, lipid peroxidation might lead to liquid ordered-liquid disordered phase separation in membranes [23], and this might favor pore formation. Thus, it would be interesting to account for phase-separated membranes (or lipid rafts) as well, as they can play a crucial role in cell permeability. Specifically, the interface region between these rafts (e.g., gel and fluid domains), the thickness mismatch between both phases, the membrane elasticity and the lipid packing, all together can affect the membrane permeability [24].

Likewise, studies on LTP-induced oxidation of certain proteins (e.g., hEGF; see section 1.2(c)) typically account for only some effects, but in reality, many more proteins play a role in plasma medicine. Specifically, for plasma cancer immunotherapy, it would be important to investigate the interaction between immune cell proteins and cancer cell proteins, under native conditions and oxidative stress (as induced by LTP). Protein-protein interactions have indeed been computationally studied in many different fields (e.g., [25]). While the effect of mutations in proteins and drugs that block the binding of immune checkpoint proteins to receptors has been simulated (e.g., [26]), the interaction between immune cell and cancer cell proteins under oxidative stress, as induced by LTP, has only been the subject of one recent study [27] (see also next paragraph), showing the need for such type of modeling, for a better insight in plasma cancer immunotherapy.

In general, however, molecular-level simulations certainly contribute to a better understanding in plasma medicine, and there are some recent examples in literature on the combined *in silico, in vitro* and even *in vivo* studies for the effect of LTP on specific biomolecules, where simulations did provide a better understanding of the experiments. For instance, Lin *et al.* [27] performed a combined *in silico, in vitro* (with 3D tumor models) and *in vivo* study on the effect of LTP on CD47 (a key innate immune checkpoint, which was found to be modulated after LTP treatment). Simulations revealed that the potential oxidized salt-bridges are responsible for conformational changes. US simulations of CD47 demonstrated that oxidation-induced conformational changes reduce its binding affinity with its receptor (signal-regulatory protein alpha; SIRPα), thus providing new insight into the potential of LTP for cancer immunotherapy, i.e., for reducing immunosuppressive signals on the surface of cancer cells. Also, Shaw *et al.* [28] used MD simulations to support *in vitro* and *in ovo* experiments on the synergistic cytotoxic effect of PTL with melittin (a small peptide component of bee venom, with reported anti-cancer effects *in vitro* and *in vivo*). Indeed, the simulations revealed that LTP-induced oxidation of the PLB results in a lower energy barrier for translocation of melittin compared to the non-oxidized PLB. Likewise, in [29] the molecular effect of LTP-induced oxidative stress on the interaction between CD44 and hyaluronan (HA) was investigated. Multi-level atomistic simulations revealed a drop in the binding free energy of HA to CD44 upon oxidation, supporting the experimental results, and the hypothesis that LTP-induced oxidation disturbs the CD44–HA interaction, which can inhibit proliferative signaling pathways inside tumor cells, and induce cancer cell death.

Finally, besides the molecular-level simulations, it would be of great interest to study cell signaling pathways, which might also explain the selectivity of plasma for cancer therapy [30]. Indeed, experiments indicated that two apoptosis-inducing signaling pathways, that are both inhibited by catalase in the extracellular compartment, could be reactivated upon inactivation of catalase in the cell membrane of cancer cells. If LTP could inactivate catalase, it may induce cancer cell death upon reactivating these signaling pathways. It is clear that a better understanding of this apoptosis-inducing mechanism and the role of catalase inactivation would be of great value. Such mechanisms can obviously not be simulated by molecular-level simulations, and macroscopic reaction kinetics modeling would be needed to study the mechanisms of the cell's antioxidant defense system and redox signaling, as extensively used in the field of redox biology (e.g., [31]). Developing such models is very challenging, and would have to be done step-by-step, starting with simple model systems, and gradually adding complexity. This has recently been demonstrated by Bengtson and Bogaerts, who developed a mathematical model to study the role of catalase inactivation for reactivating the two above-mentioned apoptotic signaling pathways, trying to understand the underlying cause of the anti-cancer effect of LTP [32]. In addition, they also developed another model to describe the key species and features of the cellular response, in an attempt to quantify the (possible) selective and synergistic anti-cancer effects of PTL. The latter model investigated the cell susceptibility towards exogenous $H_2O_2$ (alone and in combination with $NO_2^-$), in terms of two variables, i.e., the $H_2O_2$ membrane diffusion rate constant and the intracellular catalase concentration, and predicted that the maximum intracellular $H_2O_2$ concentration could be used to quantify the cell susceptibility towards exogenous $H_2O_2$ [33]. However, clearly more modeling efforts in this field are needed, to gain further insight in the underlying mechanisms of the (selective) anti-cancer action of LTP.

In general, it is clear that much more research will be needed to elucidate all the underlying mechanisms of LTP for biomedical applications. However, this section showed that computer simulations can definitely contribute to elucidate some pieces of the puzzle.

* AB and AY acknowledge financial support from the Research Foundation – Flanders (FWO; grant number 1200219N).

## IX. CONCLUDING REMARKS

On the fundamental scientific level, the advent of the application of low temperature plasma in biology and medicine has allowed for the creation of new scientific knowledge regarding the interaction of the 4th state of matter with soft matter, such as biological cells and tissues. Twenty-five years ago, such knowledge was simply missing. For decades plasma technology has been playing a major role in several industries that serve our modern society, but its introduction to the biomedical field is comparatively recent. Nonetheless, the potential socio-economic impact of using atmospheric pressure low temperature plasma to overcome outstanding hygiene, safety, and medical challenges cannot be overstated. As shown in this paper, LTP is today a well-established and accepted technology that is helping solve many serious issues related to public health, food and water safety, and others. However, to push the field to a higher level of maturity, more fundamental scientific work and novel advanced application concepts are still needed. To help facilitate this process, the team of experts who contributed to this roadmap paper have proposed viable ideas and potential approaches that should carry the field forward into the next decade.